\documentclass[twocolumn,showpacs,preprintnumbers,amsmath,amssymb]{revtex4-1}

\usepackage{graphicx}
\usepackage{dcolumn}
\usepackage{bm}
\usepackage{amsmath}
\usepackage{amssymb}
\usepackage{multirow}

\DeclareMathOperator{\sgn}{sgn}
\DeclareMathOperator{\erf}{erf}
\DeclareMathOperator{\erfc}{erfc}

\begin{document}

\preprint{}

\title{Long optical lattice vibrations and dielectric function of monolayer hexagonal boron nitride}

\author{J.-Z. Zhang}
 \email{phyjzzhang@jlu.edu.cn}
 \affiliation{School of Physics, Jilin University, Changchun, 130012, China.
}

\date{\today}

\begin{abstract}
Using two pairs of lattice equations [resembling Huang's equations for bulk crystals, Proc. Roy. Soc. A {\bf 208}, 352 (1951)] deduced from a microscopic dipole lattice model taking into account electronic polarization (EP) of ions and local field effects (LFEs) self-consistently, in-plane and out-of-plane optical vibrations in two-dimensional (2D) hexagonal BN are studied theoretically. 
The three mutually independent coefficients of either pair of lattice equations are determined by a set of three generally known quantities such as the 2D  electronic and static susceptibilities and phonon frequency, making the lattice equations very useful for calculating the lattice dynamical properties.  
Explicit expressions are obtained for lattice vibrational energy density, and phonon dispersion, group velocity and density of states.   
The transparent phonon dispersion relations describe the previous numerical calculations very well, and the longitudinal optical (LO) phonon dispersion relation is identical to the analytical expression of Sohier {\it et al}. [Nano Lett. {\bf 17}, 3758 (2017)], and it expresses the degeneracy of the LO and transverse optical (TO) modes at $\Gamma$ and their splitting at finite wavevectors due to the long-range macroscopic field. The out-of-plane phonon frequency is finite owing to ionic EP.   
A 2D lattice dielectric function $\epsilon(k,\omega)$ is derived--due solely to the LO vibrations--which also allows the LO phonon dispersion to be rederived simply from $\epsilon(k,\omega)=0$, similar to the bulk case.   
A 2D Lyddane--Sachs--Teller relation  and a frequency--susceptibility relation are obtained for the in-plane and out-of-plane vibrations, respectively, connecting the phonon frequencies to the 2D dielectric functions or susceptibilities.   
Using three first-principles calculated parameters, 
the lattice dynamical properties are studied comprehensively, particular attention being paid to the EP and LFEs.  The ionic EP and LFEs should be included simultaneously, but otherwise neglecting either or both causes large discrepancies to the calculated dynamical properties. 
The rigid-ion model cannot properly describe the optical vibrations, which, for instance, yields a 15\%-19\% larger phonon frequency and 77\% smaller Born charge for in-plane motion, and an infinitely large phonon frequency and four times larger Born charge for out-of-plane motion. 
With no LFEs or EP, the LO modes display very small linear dispersion, nearly flat in the long wavelength region, which is distinct from the LO phonon dispersion calculated after including both LFEs and EP. 
Furthermore, with ionic EP included, the LFEs increase the 2D susceptibility and Born charge by two and three times for in-plane vibrations but 
reduce both significantly for out-of-plane vibrations.   
\end{abstract}
\maketitle

\section{Introduction}

Since the discovery of graphene two-dimensional (2D) materials have become a subject of intense research due to their novel mechanical, electronic and optical properties   \cite{Geim:2013}. 
Monolayer (ML) hexagonal Boron nitride (hBN) and transition metal dichalcogenides (TMDs) such as MoS$_2$ are two types of prominent 2D binary crystals with a similar honeycomb lattice structure to graphene. While graphene is a purely covalent material, ML hBN is a 2D polar crystal with mixed covalent and ionic bonding. Also distinct from graphene, a semimetal, 2D hBN is an insulator with a large bandgap  $\sim$7 eV \cite{Rasmussen:2016,Sahin:2009}. Further, piezoelectricity occurs in ML hBN owing to inversion symmetry breaking. In recent years ML hBN has attracted intense interest due to its strong piezoelectric response \cite{Naumov:2009,Droth:2016,Michel:2017}, large in-plane stiffness and strong resistance to stretching  \cite{Sahin:2009,Songl:2010,Andrew:2012,Thomas:2016}, and also its capability to combine with graphene due to lattice match as a key material in layered graphene/hBN heterostructure electronic devices \cite{Ponomarenko:2011,Haigh:2012}.

Optical phonons are key scattering partners of electrons  \cite{Ridley:2013} and play crucial roles in carrier transport in  electronic devices. In a polar crystal 
long optical lattice vibrations are closely connected to the electric fields  \cite{Born:1954}. While the {\it macroscopic} electric field $\mathbf{E}$ is an average field, with the averaging being made to the total field of all the ions over a lattice cell \cite{Born:1954,Kittel:2004}, the {\it local} field on any particular ion $\mathbf{E}_l$, as was introduced by Lorentz, is the total electric field after deduction of the contribution due to the ion itself. In a {\it bulk} polar crystal the Lorentz relation connects the local and macroscopic fields with the {\it macroscopic}  dielectric polarization $\mathbf{P}$, $\mathbf{E}_l=\mathbf{E}+4\pi\mathbf{P}/3$  \cite{Born:1954,Kittel:2004}, where the last term is the {\it inner} field \cite{Born:1954}.  The electric field acting on each ion of the crystal polarizes it by inducing an electric dipole moment, thus causing electronic polarization (EP) to the ion, or concisely ionic polarization.  A treatment neglecting  (including) such EP is termed rigid (polarizable) ion model, i.e., RIM and PIM.  
Using the Lorentz relation whilst considering the ions to be polarizable with a microscopic model, Huang deduced a pair of equations describing the long optical vibrations of isotropic bulk crystals, $\ddot{\mathbf{w}}=b_{11}\mathbf{w}+b_{12}\mathbf{E}$,  $\mathbf{P}=b_{21}\mathbf{w}+b_{22}\mathbf{E}$ ($b_{12}=b_{21}$) \cite{Huang:1951,Huang:1951a,Born:1954}, where $\mathbf{w}$ describes the optical displacement of the unit cell, 
$\mathbf{w}=\sqrt{\bar{m}/v_a}(\mathbf{u}_1-\mathbf{u}_2)$, $\mathbf{u}_1$ and $\mathbf{u}_2$ being the displacements of the positive and negative ions, $\bar{m}$ being their reduced mass and $v_a$ the cell volume. 
The lattice dielectric function can be deduced directly from Huang's equations, allowing one to conveniently express the $b$-coefficients in terms of experimentally measurable quantities such as the static and high-frequency dielectric constants $\epsilon_0$ $\epsilon_{\infty}$ and infrared dispersion frequency $\omega_0$ \cite{Born:1954}. Solving Huang's equations and the equation of electrostatics yields the  longitudinal and the transverse optical (LO and TO) modes, with TO phonon frequency $\omega_t$ equal to  $\omega_0$. Further, 
there is a macroscopic electrostatic field associated with the LO modes making their  frequency $\omega_l$ higher than $\omega_t$, with the frequency ratio given by 
 the Lyddane--Sachs--Teller (LST) relation  \cite{Lyddane:1941}, $\omega_l/\omega_t=\sqrt{\epsilon_0/\epsilon_{\infty}}$, which can also be rederived from Huang's equations \cite{Born:1954}.

In experimental study \cite{Rokuta:1997}, phonon spectra for ML hBN on Ni and Pt  have been measured by electron energy loss spectroscopy. 
Phonon spectra of ML hBN are usually calculated by diagonalizing the dynamical matrix  which is obtained by first-principles calculation, or a tight-binding or an empirical model.  Phonon spectra have been studied for ML hBN within the local-density approximation in density functional theory (DFT) \cite{Miyamoto:1995,Illera:2017}. As a ML of hBN can be rolled up to form a BN nanotube, first-principles \cite{Ferrabone:2011} and tight-binding \cite{Sanchez:2002} calculations have been performed to make a comparative study of dielectric polarizabilities  \cite{Ferrabone:2011} as well as phonon spectra \cite{Sanchez:2002} of hBN MLs and nanotubes. Phonon modes of 2D hBN calculated from first principles \cite{Serrano:2007,Topsakal:2009,Sahin:2009,Michel:2017,Sohier:2017}  or an empirical force constant model \cite{Michel:2009,Michel:2011,Michel:2012} have been compared with those of three-dimensional (3D) bulk hBN \cite{Serrano:2007,Topsakal:2009,Michel:2011,Michel:2012} or other 2D honeycomb materials such as TMDs \cite{Michel:2017,Sohier:2017} and group III nitrides \cite{Sahin:2009}. In these studies, there are several common features present in the phonon spectra for ML hBN \cite{Sanchez:2002,Serrano:2007,Topsakal:2009,Sahin:2009,Michel:2009,Sohier:2017}: (i) the LO and TO modes are degenerate at the $\Gamma$ point but split up at a finite wavevector with the LO mode  having a higher frequency; (ii) overbending occurs in the LO phonon dispersion curve so the LO modes have maximum frequency not at $\Gamma$, but at an intermediate point away from the Brillouin zone edges;  (iii) both TO and ZO (i.e., out-of-plane optical) modes show a nondispersive character at long wavelengths with a nearly constant frequency. While most of these studies are performed by numerical methods, only few studies have used analytical approaches \cite{Sanchez:2002,Michel:2009,Sohier:2017}. 
The degeneracy has been proved to be due to the macroscopic field's in-plane component  vanishing at zero wavevector \cite{Sanchez:2002}, and analytical expressions have been obtained later for the long-wavelength dispersion of LO and TO modes \cite{Michel:2009,Sohier:2017}.  In Ref.\cite{Sohier:2017} an LO phonon dispersion relation is  derived using a simple model, in which the relationship between the squared LO and TO phonon frequencies for bulk materials \cite{Cochran:1962,Giannozzi:1991} is generalized and used for the 2D materials, and the parameters in the dispersion relation are obtained from their first-principles calculation so both ionic EP and LFEs are taken into account. 
The analytical theory of Michel {\it et al}. \cite{Michel:2009} tackles the dynamical matrix, which is obtained based on a microscopic RIM, to find the vibrational modes; while the  eigensolutions of the dynamical matrix are numerically calculated for an arbitrary wavevector, the analytical phonon dispersions are obtained for small wavevectors. 
In the study \cite{Michel:2009} ionic polarization has not been accounted for, as the inclusion of EP presents a challenge and makes it more difficult to obtain analytical solutions to the polar optical vibrations. Equally important are the {\it local fields} on the ions, which have been found to be very strong in 2D hBN \cite{Mikhailov:2013}. Further, the local fields and EP are interdependent \cite{Born:1954} and should be included in a self-consistent manner, and therefore the addition of local field effects (LFEs) makes the lattice-dynamical solutions for a general wavevector more complicated. For the long wavelengths, however one expects that the 2D lattice motion can be described on a macroscopic basis, i.e., using macroscopic quantities such as the macroscopic field and dielectric polarization, by lattice equations like Huang's equations for bulk crystals. Such equations need to be deduced from a microscopic model so as to include the intricate EP and LFEs. 
So far as we know, there are no equations of motion for the macroscopic description of the 2D lattice vibrations, and ionic EP or LFEs on the lattice dynamical properties of the 2D crystals have not been studied  systematically.  From the viewpoint of basic research, ML hBN, the structurally simplest 2D polar crystal, provides a model system for analytically studying the 2D optical vibrations and further local field and polarizable ion effects.

In this paper, we study long wavelength optical lattice vibrations and local field and polarizable ion effects for ML hBN with an analytical approach. For this purpose, 
we deduce lattice equations for optical vibrations using a microscopic model that includes the ionic EP and LFEs. We make the deduction with Huang's approach, by introducing the {\it macroscopic field} into the equation of motion whilst constructing an equation for the {\it macroscopic} dielectric polarization by adding up the two contributions due to the lattice displacement and the induced electric polarization. We solve the simultaneous lattice equations and equation of electrostatics, rather than solve the dynamical matrix equations as in previous studies, to obtain  explicit expressions for the optical modes, which can describe the key features of the 2D  phonon modes. 
We derive a 2D longitudinal lattice dielectric function $\epsilon(k,\omega)$ which also allows one to rederive the LO phonon dispersion simply from $\epsilon(k,\omega)=0$.   
We also deduce a LST relation, a 2D counterpart of the LST relation in bulk, for in-plane motion and  a frequency--susceptibility relation for out-of-plane motion.    
Using first-principles calculated quantities  
we study the lattice dynamical properties for 2D hBN and discuss in great detail the  local field and polarizable ion effects. 

This paper is organized as follows. In Section II, a deduction of 2D lattice equations for in-plane and out-of-plane motion is made, through a 2D Lorentz relation, from a microscopic dipole lattice model including LFEs and EP. From the lattice equations, the dispersion relations of the optical modes are deduced, followed by a derivation of the phonon group velocity and density of states (DOS).  Then the dynamical lattice dielectric susceptibilities are derived yielding 
the relations relating the static and electronic susceptibilities to the $a_{12}$ and $c_{12}$ coefficients of the lattice equations. Further, a 2D longitudinal lattice dielectric function is deduced after considering the general test charge distributions, and from it a 2D LST relation follows. 
In Section III, we present results of the in-plane and out-of-plane optical vibrations in ML BN. A comparison of microscopic quantities such as effective charges and spring force constants is made, which are calculated when knowing three quantities including the  electronic and static susceptibilities from independent first-principles calculations.  
Then various lattice-dynamical quantities, such as the Born charge, the phonon  dispersion and the static and electronic susceptibilities, are compared with those obtained from a RIM, with or without LFEs taken into account, to study the EP and LFEs on the lattice dynamical properties. Finally, Section V summarizes the main results obtained. In Appendix A, we show that using the 2D Clausius-Mossotti relation obtained, 
 the unit-cell atomic polarizability falls in an interval for in-plane or out-of-plane  polarization, which is used to evaluate the LFEs on the phonon dispersion and 2D dielectric susceptibilities.    
In Appendix B we show that in terms of macroscopic theory the relations $a_{12}=a_{21}$ and $c_{12}=c_{21}$, which 
 connect the coefficients of the lattice equations,  
follow from the principle of energy conservation. Further we obtain a lattice-vibrational energy density as a function of the optical displacement and electric field for in-plane or out-of-plane vibrations.

\section{Theory}

\subsection{Equations of motion and lattice polarization}

ML hBN, a 2D binary crystal with point group D$_{3h}$, is composed of two sublattices of B and N (as shown in Fig.~\ref{fig1}), labeled with $\kappa=1, 2$, respectively.  Let $m_{\kappa}$ and $e_{\kappa}$ be the mass and charge of the type $\kappa$ ions , and let $e_1=-e_2=e_a$, where $e_a$ is the static effective charge \cite{Karch:1997} due to electron charge transfer $-e_a$ from B to N in 2D h-BN, $e_a>0$ \cite{Topsakal:2009,Michel:2017}. 
The masses of the B and N atoms are $m_1$=10.81 Da and $m_2$=14.01 Da. 
In the dipole lattice model \cite{Born:1954} each ion site of type $\kappa$ is occupied by an electric dipole $\mathbf{p}_{\kappa}$ which arises due partly to the ionic displacement $\mathbf{u}_{\kappa}$ and partly to the induced electric moment $\boldsymbol{\mu}_{\kappa}$ on the ion. 
Associated with the long wavelength optical modes, there is macroscopic dielectric polarization $\mathbf{P}=(\mathbf{p}/s)e^{i\mathbf{k}\cdot\boldsymbol{\rho}}\delta(z)$, where $\mathbf{p}$ is the total dipole moment $\mathbf{p}=\mathbf{p}_1+\mathbf{p}_2$  of a unit cell with area $s=\sqrt{3}a^2/2$, $a$ being the lattice constant $a$=2.5 $\AA$ (see Fig.~\ref{fig1}). The $\delta$ function describes the  dependence of the polarization on $z$ for a ML in the plane $z=0$. $\mathbf{k}$ is the 2D wave vector, and $\boldsymbol{\rho}=(x,y)$ is the position vector in the  plane parallel to the ML. 

We first consider in-plane optical vibrations; that is, the displacements $\mathbf{u}_{\kappa}$ and dipole moments $\mathbf{p}_{\kappa}$ lie in the ML plane. The macroscopic electric field $\mathbf{E}$ due to the charge density $-\nabla\cdot\mathbf{P}$ is given by the equation of electrostatics $\nabla\cdot(\mathbf{E}+4\pi\mathbf{P})=0$, 
the electric field $\mathbf{E}$ being an irrotational field, $\mathbf{E}=-\nabla\phi$ [$\phi$ is the electrostatic potential, $\phi(\boldsymbol{\rho},z)=\varphi(z)e^{i\mathbf{k}\cdot\boldsymbol{\rho}}$]. To solve the corresponding Poisson's equation 
$\nabla^2\phi(\boldsymbol{\rho},z)=4\pi i\mathbf{p}\cdot \mathbf{k}e^{i\mathbf{k}\cdot\boldsymbol{\rho}}\delta(z)/s$, we expand $\varphi(z)$ and $\delta(z)$,  
\begin{equation}
\varphi(z)=\int_{-\infty}^{\infty}\hat{\varphi}(q)e^{iqz}dq,
\label{vphx}
\end{equation}
\begin{equation}
\delta(z)=\frac{1}{2\pi}\int_{-\infty}^{\infty}e^{iqz}dq. 
\label{delx}
\end{equation}

We find $\hat{\varphi}(q)=-2i\mathbf{p}\cdot \mathbf{k}/[s(k^2+q^2)]$ and from Eq.~(\ref{vphx}) $\varphi(z)=-2\pi i\mathbf{p}\cdot\mathbf{k}e^{-k\lvert z\rvert}/(sk)$, and  
 then obtain for the macroscopic field its components in the directions parallel and perpendicular to the ML:     
\begin{subequations} 
\begin{equation}
\mathbf{E}_{\boldsymbol{\rho}}(\boldsymbol{\rho},z)=-\frac{2\pi}{s}\frac{\mathbf{k}}{k}~\mathbf{p}\cdot\mathbf{k}e^{-k\lvert z\rvert}e^{i\mathbf{k}\cdot\boldsymbol{\rho}}, 
\label{Ero1}
\end{equation}
\begin{equation}
\mathbf{E}_z(\boldsymbol{\rho},z)=-\mathbf{e}_z\frac{2\pi i}{s}\mathbf{p}\cdot\mathbf{k}~\sgn(z)e^{-k\lvert z\rvert}e^{i\mathbf{k}\cdot\boldsymbol{\rho}}.  
\label{Ez1}
\end{equation}
\end{subequations} 
Here $\sgn(z)$ is the sign function and in particular the $z$-component of the field is zero in the ML.    

Recall that in bulk polar crystals the macroscopic field is strongly nonanalytical close to zero wavevector, i.e., its limiting value depending on the direction along which zero wavevector is approached \cite{Born:1954}, and the LO vibrations have a finite field at small wavevectors. Distinct from the field in 3D crystals, the macroscopic field [Eqs.~(\ref{Ero1}) and (\ref{Ez1})] of 2D hBN vanishes at a very  small wavevector for in-plane motion \cite{Sanchez:2002}, independent of the direction of the wavevector.

In a 2D dipole lattice a dipole $\mathbf{p}$ on the ion at a lattice point $\boldsymbol{\rho}_i=(x_i,y_i)$ gives rise to a field at the origin equal to $-\mathbf{p}/\rho_i^3+3\mathbf{p}\cdot\boldsymbol{\rho}_i\boldsymbol{\rho}_i/\rho_i^5$. The Lorentz local field $\mathbf{E}_{l}$, also called the exciting field \cite{Born:1954}, is the electric field acting on an ion due to all the {\it other} dipoles oscillating in the lattice, and can be written as the macroscopic field $\mathbf{E}$ plus another field $\mathbf{E}_{in}$ (namely, the {\it inner} field) \cite{Born:1954},  
$\mathbf{E}_{l}=\mathbf{E}+\mathbf{E}_{in}$. 
In a {\it long wavelength} lattice wave, the local fields at the B and N lattice points are given by the following expressions \cite{Mikhailov:2013}, respectively, 
\begin{subequations} 
\begin{equation}
\mathbf{E}_{l,1}=\mathbf{E}+Q_0~\mathbf{p}_1+Q_1~\mathbf{p}_2, 
\label{Ex1}
\end{equation}
\begin{equation}
\mathbf{E}_{l,2}=\mathbf{E}+Q_1~\mathbf{p}_1+Q_0~\mathbf{p}_2. 
\label{Ex2}
\end{equation}
\end{subequations} 
with the coefficients $Q_0$ and $Q_1$  \cite{Mikhailov:2013,Della:2016}   
\begin{subequations} 
\begin{equation}
Q_0=\sum_{(m,n)\neq (0,0)}\frac{1}{2(m^2+n^2+mn)^{3/2}a^3}\approx \frac{5.5171}{a^3}~, 
\label{Q0}
\end{equation}
\begin{equation}
Q_1=\sum_{m,n}\frac{1}{2(m^2+n^2+mn+n+1/3)^{3/2}a^3}\approx \frac{11.5753}{a^3}~. 
\label{Q1}
\end{equation}
\end{subequations} 
Let dipole $\mathbf{p}_1$ or equally $\mathbf{p}_2$ have 0.01 $e\AA$, for instance, corresponding to a displacement of 0.01 $\AA$ of the ions with charge $e$, then from these expressions one finds very strong local fields with a magnitude 1000 kV/cm \cite{Mikhailov:2013}.     

The LFEs are included through coefficients $Q_0$ and $Q_1$. Note that $\mathbf{E}$ is the macroscopic field [Eqs.~(\ref{Ero1}) and (\ref{Ez1})] rather than an external field in a simple sense \cite{Mikhailov:2013}. 
Such a simple relation between the local field and the macroscopic field is valid only for the long lattice waves, while the general expression for an arbitrary wavelength is quite complicated with both coefficients $Q_0$ and $Q_1$ dependent on the wavevector.   
In 3D iostropic polar crystals \cite{Born:1954} the difference between the local and macroscopic fields is proportional to the {\it macroscopic polarization} directly (the Lorentz relation), $\mathbf{E}_l=\mathbf{E}+4\pi\mathbf{P}/3$, as the $Q$ coefficients are equal to $4\pi/(3v_a)$ at $\mathbf{k}=0$ \cite{Born:1954}.  When approximating $Q_0=Q_1$, Eqs.~(\ref{Ex1}) and (\ref{Ex2}) can be transformed into a simple Lorentz  relation involving macroscopic {\it areal polarization} $(\mathbf{p}_1+\mathbf{p}_2)/s$, and also a Clausius-Mossotti relation can be deduced for 2D BN (Appendix A).  Expressions (\ref{Ex1}) and (\ref{Ex2}), the Lorentz relations for 2D BN, show that the finite local fields occur in ML hBN at $\mathbf{k}=0$, different from those in the 3D polar crystals where local fields vanish in the long wavelength limit \cite{Born:1954}. 

Apart from the macroscopic field and local fields, we also need to find the field change at the center of an ion of type $\kappa$ owing to its own displacement $\mathbf{u}_{\kappa}$ \cite{Born:1954}. The field is evidently equal to the field created at the ion $\kappa$ site by displacing all other ions by $-\mathbf{u}_{\kappa}$. Hence it is equal to the local  field at that ion site in a dipole lattice with displacement dipoles $\mathbf{p}_{\kappa'}=-e_{\kappa'}\mathbf{u}_{\kappa}$, where type $\kappa'=1, 2$. Substituting this dipole expression into Eqs.~(\ref{Ex1}) and (\ref{Ex2}) and putting $\mathbf{E}=0$ as wavevector $\mathbf{k}=0$, we find the field changes at the centers of the B and N ions due to their own displacements respectively,  
\begin{subequations} 
\begin{equation}
\mathbf{E}_{u,1}=-\mathbf{u}_1(e_1Q_0+e_2Q_1), 
\label{Eu1}
\end{equation}
\begin{equation}
\mathbf{E}_{u,2}=-\mathbf{u}_2(e_1Q_1+e_2Q_0).   
\label{Eu2}
\end{equation}
\end{subequations} 
The {\it total} Coulomb fields $\mathbf{E}_1$ and $\mathbf{E}_2$ at the centers of the B and N ions are simply the sums of $\mathbf{E}_{l,1}$ and $\mathbf{E}_{u,1}$  [Eqs.~(\ref{Ex1}) and (\ref{Eu1})], and $\mathbf{E}_{l,2}$ and $\mathbf{E}_{u,2}$ [Eqs.~(\ref{Ex2}) and (\ref{Eu2})], respectively, 
\begin{subequations} 
\begin{equation}
\mathbf{E}_1=\mathbf{E}+Q_0~\mathbf{p}_1+Q_1~\mathbf{p}_2+e_a(Q_1-Q_0)\mathbf{u}_1,  
\label{Eto1}
\end{equation}
\begin{equation}
\mathbf{E}_2=\mathbf{E}+Q_1~\mathbf{p}_1+Q_0~\mathbf{p}_2-e_a(Q_1-Q_0)\mathbf{u}_2,  
\label{Eto2}
\end{equation}
\end{subequations} 
where all the vectors are in the layer plane. 

Assuming that the {\it electronic} polarization of an ion is equivalent to a point-dipole  
\cite{Born:1954}, the electronic (i.e., {\it induced}) dipole moment of the ion $\kappa$ is then given by $\boldsymbol{\mu}_{\kappa}=\alpha_{\kappa}\mathbf{E}_{\kappa}$, where $\alpha_{\kappa}$ is the in-plane {\it electronic} polarizability of the ion $\kappa$.  
Then the total dipole moments on the B and N ions are 
\begin{subequations} 
\begin{equation}
\mathbf{p}_1=e_a\mathbf{u}_1+\alpha_1~\mathbf{E}_1, 
\label{pto1}
\end{equation}
\begin{equation}
\mathbf{p}_2=-e_a\mathbf{u}_2+\alpha_2~\mathbf{E}_2. 
\label{pto2}
\end{equation}
\end{subequations} 

Inserting the expressions for the total fields $\mathbf{E}_1$ and $\mathbf{E}_2$ into Eqs.~(\ref{pto1}) and (\ref{pto2}) and then rearranging the terms, we find 
\begin{subequations} 
\begin{equation}
(1-\alpha_1Q_0)\mathbf{p}_1-\alpha_1Q_1\mathbf{p}_2=e_a[1+\alpha_1(Q_1-Q_0)]\mathbf{u}_1+\alpha_1~\mathbf{E}, 
\label{pto1a}
\end{equation}
\begin{equation}
-\alpha_2Q_1\mathbf{p}_1+(1-\alpha_2Q_0)\mathbf{p}_2=-e_a[1+\alpha_2(Q_1-Q_0)]\mathbf{u}_2+\alpha_2~\mathbf{E}. 
\label{pto2a}
\end{equation}
\end{subequations}

Solving Eqs.~(\ref{pto1a}) and (\ref{pto2a}) then we can express $\mathbf{p}_1$ and $\mathbf{p}_2$ in terms of $\mathbf{u}_1$, $\mathbf{u}_2$ and $\mathbf{E}$ as follows: 
\begin{subequations} 
\begin{align}
\mathbf{p}_1&=\frac{1}{D}\Big \{e_a(1-\alpha_2Q_0)\big[1+\alpha_1(Q_1-Q_0)\big]\mathbf{u}_1 \Big. 
\nonumber \\ 
&\qquad {} -e_a\alpha_1Q_1\big[1+\alpha_2(Q_1-Q_0)\big]\mathbf{u}_2
\nonumber \\ 
&\qquad {}  \Big. +\alpha_1\big[1+\alpha_2(Q_1-Q_0)\big]\mathbf{E} \Big \}, 
\label{pt1}
\end{align}
\begin{align}
\mathbf{p}_2&=\frac{1}{D}\Big \{e_a\alpha_2Q_1\big[1+\alpha_1(Q_1-Q_0)\big]\mathbf{u}_1 \Big. 
\nonumber \\ 
&\qquad {}   -e_a(1-\alpha_1Q_0)\big[1+\alpha_2(Q_1-Q_0)\big]\mathbf{u}_2
\nonumber \\ 
&\qquad {}  \Big. +\alpha_2\big[1+\alpha_1(Q_1-Q_0)\big]\mathbf{E} \Big \},    
\label{pt2}
\end{align}
\end{subequations} 
where 
\begin{equation}
D=1-(\alpha_1+\alpha_2)Q_0-\alpha_1\alpha_2(Q_1^2-Q_0^2).  
\label{bigD}
\end{equation}

Define the areal polarization $\boldsymbol{\mathcal{P}}$ 
\begin{equation}
\boldsymbol{\mathcal{P}}=(\mathbf{p}_1+\mathbf{p}_2)/s, 
\label{defP}
\end{equation}
and introduce the optical displacement $\mathbf{w}$ 
\begin{equation}
\mathbf{w}=\sqrt{\frac{\bar{m}}{s}}(\mathbf{u}_1-\mathbf{u}_2),    
\label{smw}
\end{equation}
where $\bar{m}$ is the reduced mass, $\bar{m}=m_1m_2/(m_1+m_2)$. 

When expressions~(\ref{pt1}) and (\ref{pt2}) are substituted for $\mathbf{p}_1$ and $\mathbf{p}_2$, we obtain 
\begin{equation}
\boldsymbol{\mathcal{P}}=a_{21}\mathbf{w}+a_{22}\mathbf{E},   
\label{bigP1}
\end{equation}
where 
\begin{subequations} 
\begin{equation}
a_{21}=\frac{e_a}{D\sqrt{\bar{m}s}}\big[1+\alpha_1(Q_1-Q_0)\big]\big[1+\alpha_2(Q_1-Q_0)\big], 
\label{a21}
\end{equation}
\begin{equation}
a_{22}=\frac{1}{sD}\big[(\alpha_1+\alpha_2)+2\alpha_1\alpha_2(Q_1-Q_0)\big].   
\label{a22}
\end{equation}
\end{subequations} 
Eq.~(\ref{bigP1}) shows that the {\it macroscopic} quantity $\boldsymbol{\mathcal{P}}$ of the 2D crystal, which has a clear physical meaning as given by expressions~(\ref{defP}),  
is simplified to a sum of two contributions, one due to the optical displacement and the other due to the {\it macroscopic} field.

It is evident from Eq.~(\ref{bigP1}) that $a_{22}$ is the in-plane component $\chi_e$  of the {\it electronic} susceptibility of the 2D crystal, 
\begin{equation}
a_{22}=\chi_e.    
\label{a22alf}
\end{equation} 

When the Born charge \cite{Gonze:1997} 
\begin{equation}
e_B=\frac{e_a}{D}\big[1+\alpha_1(Q_1-Q_0)\big]\big[1+\alpha_2(Q_1-Q_0)\big], 
\label{eB}
\end{equation}
is introduced, then the coefficient $a_{21}$ relates simply to the Born charge $e_B$ by 
\begin{equation}
a_{21}=\frac{e_B}{\sqrt{\bar{m}s}}.   
\label{a21b}
\end{equation}

From Eqs.~(\ref{a22}), (\ref{a22alf}) and (\ref{eB}) we find  
\begin{equation}
a_{22}=\chi_e=\frac{1}{sQ_1}\left(\frac{e_B}{e_a}-1\right), 
\label{a22eB}
\end{equation}
showing that apart from $a_{21}$, the coefficient $a_{22}$ is also related to the Born charge. $e_B \neq e_a$ owing to the electronic polarization of the ions, and further  considering $Q_1 >0$, $e_B$ is greater than $e_a$. 

Inserting expressions (\ref{pt1}) and (\ref{pt2}) for $\mathbf{p}_1$ and $\mathbf{p}_2$ into Eqs.~(\ref{Eto1}) and (\ref{Eto2}), we express the total fields $\mathbf{E}_1$ and $\mathbf{E}_2$ in terms of the ionic displacements and macroscopic field. After  collecting like terms we can simplify the total field expressions to
\begin{subequations} 
\begin{equation}
\mathbf{E}_1=\frac{1}{D}\big[1+\alpha_2(Q_1-Q_0)\big]\big[e_aQ_1(\mathbf{u}_1-\mathbf{u}_2)+\mathbf{E}\big], 
\label{Eto1b}
\end{equation}
\begin{equation}
\mathbf{E}_2=\frac{1}{D}\big[1+\alpha_1(Q_1-Q_0)\big]\big[e_aQ_1(\mathbf{u}_1-\mathbf{u}_2)+\mathbf{E}\big].  
\label{Eto2b}
\end{equation}
\end{subequations}

The Coulomb force acting on the ion $\kappa$ consists of two parts \cite{Born:1954}, (i) the force exerted on the ionic charge $e_{\kappa}$ by the total field $\mathbf{E}_{\kappa}$, and (ii) the force exerted on the induced dipole $\boldsymbol{\mu}_{\kappa}$ by the field of all other ions. 
The latter electric force can be sought by using again the Lorentz relations  (\ref{Ex1}) and (\ref{Ex2}) as follows. Imagine that we subject the dipole $\boldsymbol{\mu}_{\kappa}$ at a particular site of ion $\kappa$ a virtual displacement $\mathbf{u}$ while keeping all other ions in their undisplaced positions. The virtual energy is then simply the interaction energy between the dipole and the field at the ion site which is created equivalently by displacing all other ions by $-\mathbf{u}$ \cite{Born:1954}, and which is thus the local field in a dipole lattice with dipole moments $\mathbf{p}_{\kappa^\prime}=-e_{\kappa^\prime}\mathbf{u}$, (type $\kappa^\prime=1, 2$). Therefore the virtual energy is 
$-\boldsymbol{\mu}_{\kappa}\cdot\mathbf{E}_{l,\kappa}$, and the force on the dipole is the negative gradient of this virtual energy with respect to virtual displacement $\mathbf{u}$, given by $\nabla _{\mathbf{u}}(\boldsymbol{\mu}_{\kappa}\cdot\mathbf{E}_{l,\kappa})$. Inserting the above dipole moment expression into Eqs.~(\ref{Ex1}) and (\ref{Ex2}) and recalling the macroscopic field $\mathbf{E}=0$, we find the local  fields and then obtain for the forces on the dipoles $\boldsymbol{\mu}_1$ and $\boldsymbol{\mu}_2$ the expressions  
$e_a(Q_1-Q_0)\boldsymbol{\mu}_1$ and $-e_a(Q_1-Q_0)\boldsymbol{\mu}_2$, respectively, where $\boldsymbol{\mu}_{\kappa}=\alpha_{\kappa}\mathbf{E}_{\kappa}$ ($\kappa=1, 2$).   

Apart from the electric forces, there are also restoring forces caused by the overlap potential between the positive and negative ions in the 2D polar crystal. The restoring forces on the B and N ions are $-K(\mathbf{u}_1-\mathbf{u}_2)$, and $-K(\mathbf{u}_2-\mathbf{u}_1)$, respectively, where the spring force constant $K$ for in-plane motion is a simple scalar rather than a tensor due to the hexagonal symmetry. 

Therefore the equations of motion for the B and N ions are given by
\begin{subequations} 
\begin{equation}
m_1\ddot{\mathbf{u}}_1=-K(\mathbf{u}_1-\mathbf{u}_2)+e_a\mathbf{E}_1+e_a\alpha_1(Q_1-Q_0)\mathbf{E}_1, 
\label{eom1}
\end{equation}
\begin{equation}
m_2\ddot{\mathbf{u}}_2=-K(\mathbf{u}_2-\mathbf{u}_1)-e_a\mathbf{E}_2-e_a\alpha_2(Q_1-Q_0)\mathbf{E}_2.  
\label{eom2}
\end{equation}
\end{subequations} 
Substitute expressions (\ref{Eto1b}) and (\ref{Eto2b}) for $\mathbf{E}_1$ and $\mathbf{E}_2$ respectively in the two equations above. Using the Born charge $e_B$ [expression (\ref{eB})]  
and introducing a force constant due to LFEs $K_e$,  
\begin{equation}
K_e=e_ae_BQ_1, 
\label{Ke}
\end{equation}
the equations of motion reduce to  
\begin{subequations} 
\begin{equation}
m_1\ddot{\mathbf{u}}_1=(K_e-K)(\mathbf{u}_1-\mathbf{u}_2)+e_B\mathbf{E}, 
\label{eom1b}
\end{equation}
\begin{equation}
m_2\ddot{\mathbf{u}}_2=(K-K_e)(\mathbf{u}_1-\mathbf{u}_2)-e_B\mathbf{E}.  
\label{eom2b}
\end{equation}
\end{subequations} 

Multiplying Eqs.~(\ref{eom1b}) and (\ref{eom2b}) by $m_2$ and $m_1$ respectively, subtracting and then dividing by $(m_1+m_2)$, we find 
\begin{equation}
\bar{m}(\ddot{\mathbf{u}}_1-\ddot{\mathbf{u}}_2)=(K_e-K)(\mathbf{u}_1-\mathbf{u}_2)+e_B\mathbf{E}.  
\label{eom12b}
\end{equation}

On expressing $(\mathbf{u}_1-\mathbf{u}_2)$ in terms of $\mathbf{w}$ [Eq.~(\ref{smw})], we obtain 
\begin{equation}
\ddot{\mathbf{w}}=a_{11}\mathbf{w}+a_{12}\mathbf{E},   
\label{eomw1}
\end{equation}
where
\begin{subequations} 
\begin{equation}
a_{11}=\frac{1}{\bar{m}}(K_e-K)=-\omega_0^2, 
\label{a11}
\end{equation}
\begin{equation}
a_{12}=\frac{e_B}{\sqrt{\bar{m}s}},
\label{a12}
\end{equation}
\end{subequations} 
$\omega_0$ being the intrinsic oscillator frequency, i.e., in the absence of macroscopic field $\mathbf{E}$. 

Comparing Eqs.~(\ref{a21b}) and (\ref{a12}) we find the relation 
\begin{equation}
a_{12}=a_{21}.  
\label{a12a21}
\end{equation}

The equation of motion (\ref{eomw1}) and the polarization equation (\ref{bigP1}) describe the in-plane polar optical vibrations of 2D BN, the frequencies of which will be derived in Sec. III below. The lattice vibrations we are dealing with are of long wavelengths, and the pair of equations (\ref{eomw1}) and (\ref{bigP1}) constitute a macroscopic description of the lattice motion. 
It is shown in Appendix B that, from the viewpoint of the {\it macroscopic} theory, the relation  (\ref{a12a21}) is due to the principle of energy conservation, and further this relation makes it possible to define an areal energy density [expression~(\ref{endenuh})] from which the lattice equations (\ref{eomw1}) and (\ref{bigP1}) can be rederived. 
It is evident from Eqs.~(\ref{eom1b}) and (\ref{eom2b}) that the center of mass of the two atoms remains stationary  (frequency $\omega=0$), yielding trivial nondynamical solutions.  
We note that for clearness $\mathbf{E}$ appearing in all the equations above is used to represent the in-plane component of the macroscopic field in the ML, i.e., $\mathbf{E}=\mathbf{E}_{\boldsymbol{\rho}}(\boldsymbol{\rho},0)$.

Now we consider ionic motion perpendicular to the layer plane, in which case the displacements $\mathbf{u}_{\kappa}$ and dipole moments $\mathbf{p}_{\kappa}$ are parallel to $\mathbf{e}_z$. 
Let $\alpha_{\kappa}^\prime$ denote the electronic polarizability of the type $\kappa$ ions; note that in general $\alpha_{\kappa}^\prime\neq\alpha_{\kappa}$ (the latter is the polarizability for in-plane motion), as they are simply components of the polarizability tensor \cite{Born:1954}. Let $K^\prime$ be the force constant associated with the perpendicular motion.
Considering the anisotropic 3D charge density distribution, another static {\it effective} charge $e_a^\prime$ exists likewise, which may differ from $e_a$ for in-plane motion (anisotropy), and is needed in the {\it point-ion} model for the perpendicular motion; thus the charges on the ions are $e_1^\prime=-e_2^\prime=e_a^\prime$. 

Solving the Poisson equation yields the electrostatic potential
\begin{equation}
\varphi(z)=2\pi \mathbf{p}\cdot\mathbf{e}_z\sgn(z)e^{-k\lvert z\rvert}/s, 
\label{Pophez}
\end{equation}
and then the $z$- and in-plane components of the macroscopic field follow,     
\begin{subequations} 
\begin{equation}
\mathbf{E}_z(\boldsymbol{\rho},z)=-\frac{4\pi}{s}\mathbf{p}\left[\delta(z)-\frac{1}{2}ke^{-k\lvert z\rvert}\right]e^{i\mathbf{k}\cdot\boldsymbol{\rho}},   
\label{Ez2}
\end{equation}
\begin{equation}
\mathbf{E}_{\boldsymbol{\rho}}(\boldsymbol{\rho},z)=-\frac{2\pi i}{s}\mathbf{k}~\mathbf{p}\cdot\mathbf{e}_z\sgn(z)e^{-k\lvert z\rvert}e^{i\mathbf{k}\cdot\boldsymbol{\rho}}.  
\label{Ero2}
\end{equation}
\end{subequations} 
We note that (i) there is a $\delta(z)$ term in the $z$-component of the macroscopic field, which reflects the microscopic character of the polarization in terms of its atomic scale when the dipoles point to the $z$ direction.  
This term follows the dielectric polarization 
$\mathbf{P}$ according to $-4\pi\mathbf{P}$, which is also true when the $\delta(z)$ function of $\mathbf{P}$ is generalized to an arbitrary function $f(z)$;  (ii) the in-plane component of the field is zero in the ML.

The local fields at the B and N sites are given by   
\begin{subequations} 
\begin{equation}
\mathbf{E}_{l,1}=\mathbf{E}+Q_0^\prime~\mathbf{p}_1+Q_1^\prime~\mathbf{p}_2, 
\label{Ex1z}
\end{equation}
\begin{equation}
\mathbf{E}_{l,2}=\mathbf{E}+Q_1^\prime~\mathbf{p}_1+Q_0^\prime~\mathbf{p}_2,  
\label{Ex2z}
\end{equation}
\end{subequations} 
respectively, where the coefficients $Q_0^\prime$ and $Q_1^\prime$ are negative, $Q_0^\prime=-2Q_0$, and $Q_1^\prime=-2Q_1$ \cite{Mikhailov:2013}. The long wavelength macroscopic field has a $\delta(z)$ form [Eq.~(\ref{Ez2})] but in calculating the field change experienced by a type $\kappa$ ion owing to its own displacement $\mathbf{u}_{\kappa}$, the contribution of this macroscopic field is zero, i.e., $4\pi\mathbf{u}_{\kappa}\sum_{\kappa^\prime}e_{\kappa^\prime}^\prime\delta(z)/s=4\pi\mathbf{u}_{\kappa}(e_1^\prime+e_2^\prime)\delta(z)/s=0$. Thus the field changes at the centers of B and N owing to their own displacements are   
\begin{subequations} 
\begin{equation}
\mathbf{E}_{u,1}=-\mathbf{u}_1(e_1^\prime Q_0^\prime+e_2^\prime Q_1^\prime), 
\label{Eu1z}
\end{equation}
\begin{equation}
\mathbf{E}_{u,2}=-\mathbf{u}_2(e_1^\prime Q_1^\prime+e_2^\prime Q_0^\prime),  
\label{Eu2z}
\end{equation}
\end{subequations} 
respectively. Similarly, in calculating the force exerted on the induced dipole $\boldsymbol{\mu}_{\kappa}$ at a type $\kappa$ ion by the field of all other ions, the 
macroscopic field makes no contribution to this force again because the net charge per cell vanishes. 
Then repeating the process as before, we find that with replacements $Q_0\rightarrow Q_0^\prime$, $Q_1\rightarrow Q_1^\prime$, $e_a\rightarrow e_a^\prime$, $\alpha_{\kappa}\rightarrow\alpha_{\kappa}^\prime$ and $ K\rightarrow K^\prime$, the equations and expressions above for the in-plane motion are applicable to the out-of-plane motion.

The Born charge is 
\begin{equation}
e_B^\prime=\frac{e_a^\prime}{D^\prime}\big[1+\alpha_1^\prime(Q_1^\prime-Q_0^\prime)\big]\big[1+\alpha_2^\prime(Q_1^\prime-Q_0^\prime)\big], 
\label{eBz}
\end{equation}
where
\begin{equation}
D^\prime=1-(\alpha_1^\prime+\alpha_2^\prime)Q_0^\prime-\alpha_1^\prime\alpha_2^\prime(Q_1^{\prime 2}-Q_0^{\prime 2}).  
\label{bigD1}
\end{equation}

The polarization equation is 
\begin{equation}
\boldsymbol{\mathcal{P}}=c_{21}\mathbf{w}+c_{22}\mathbf{E},   
\label{bigP1z}
\end{equation}
where $\mathbf{w}$ is the optical displacement as given by Eq.~(\ref{smw}) and $c_{22}$ is the out-of-plane component $\chi_e^\prime$ of the electronic susceptibility of the 2D crystal, 
\begin{equation}
c_{22}=\chi_e^\prime,     
\label{c22alf}
\end{equation} 
and $c_{21}$ and $c_{22}$ relate to the Born charge via 
\begin{subequations} 
\begin{equation}
c_{21}=\frac{e_B^\prime}{\sqrt{\bar{m}s}},  
\label{c21z}
\end{equation}
\begin{equation}
c_{22}=\chi_e^\prime=\frac{1}{sQ_1^\prime}\left(\frac{e_B^\prime}{e_a^\prime}-1\right). 
\label{c22z}
\end{equation}
\end{subequations}

The equation of motion is given by 
\begin{equation}
\ddot{\mathbf{w}}=c_{11}\mathbf{w}+c_{12}\mathbf{E},   
\label{eomw1z}
\end{equation}
where
\begin{subequations} 
\begin{equation}
c_{11}=\frac{1}{\bar{m}}(K_e^\prime-K^\prime)=-\omega_0'^2 \quad  (K_e^\prime=e_a^\prime e_B^\prime Q_1^\prime), \label{c11z}
\end{equation}
\begin{equation}
c_{12}=c_{21}=\frac{e_B^\prime}{\sqrt{\bar{m}s}}, 
\label{c12z}
\end{equation}
\end{subequations} 
$\omega_0'$ being the intrinsic oscillator frequency. 

In the lattice equations (\ref{eomw1z}) and (\ref{bigP1z}) $\mathbf{E}$ is the field in the ML, $\mathbf{E}(\boldsymbol{\rho},0)$, and evidently $\mathbf{E}(\boldsymbol{\rho},0)=\mathbf{E}_z(\boldsymbol{\rho},0)$ for the out-of-plane  vibrations. 
The areal energy density associated with the out-of-plane optical vibrations is given by expression~(\ref{endenuv}). 

When the equations for in-plane motion [Eqs.~(\ref{eomw1}) and (\ref{bigP1})] and out-of-plane motion [Eqs.~(\ref{eomw1z}) and (\ref{bigP1z})] are considered simultaneously, they can be rewritten for clarity as 
\begin{subequations} 
\begin{equation}
\ddot{\mathbf{w}}_{\boldsymbol{\rho}}(\boldsymbol{\rho})=a_{11}\mathbf{w}_{\boldsymbol{\rho}}(\boldsymbol{\rho})+a_{12}\mathbf{E}_{\boldsymbol{\rho}}(\boldsymbol{\rho},0),   
\label{eomw1co}
\end{equation}
\begin{equation}
\boldsymbol{\mathcal{P}}_{\boldsymbol{\rho}}(\boldsymbol{\rho})=a_{21}\mathbf{w}_{\boldsymbol{\rho}}(\boldsymbol{\rho})+a_{22}\mathbf{E}_{\boldsymbol{\rho}}(\boldsymbol{\rho},0),   
\label{bigP1co}
\end{equation}
\end{subequations} 
and 
\begin{subequations} 
\begin{equation}
\ddot{\mathbf{w}}_z(\boldsymbol{\rho})=c_{11}\mathbf{w}_z(\boldsymbol{\rho})+c_{12}\mathbf{E}_z(\boldsymbol{\rho},0),   
\label{eomw1zco}
\end{equation}
\begin{equation}
\boldsymbol{\mathcal{P}}_z(\boldsymbol{\rho})=c_{21}\mathbf{w}_z(\boldsymbol{\rho})+c_{22}\mathbf{E}_z(\boldsymbol{\rho},0),   
\label{bigP1zco}
\end{equation}
\end{subequations} 
respectively, where $a_{12}=a_{21}$ and $c_{12}=c_{21}$. These equations have similar forms to Huang's equations for bulk crystals \cite{Huang:1951,Born:1954}.

\subsection{In-plane and out-of-plane optical modes}

The in-plane optical vibration modes can be obtained from Eqs.~(\ref{eomw1}) and (\ref{bigP1}) in conjunction with the equation of electrostatics 
$\nabla\cdot(\mathbf{E}+4\pi\mathbf{P})=0$, where $\mathbf{P}$ is the dielectric  polarization (namely, a dipole moment per unit volume), $\mathbf{P}=\boldsymbol{\mathcal{P}}\delta(z)$, and 
 $\mathbf{E}$ is an irrotational field, $\mathbf{E}=-\nabla\phi$. Let $\mathbf{w}(\boldsymbol{\rho})=\mathbf{w}_0e^{i\mathbf{k}\cdot\boldsymbol{\rho}}$ and   electrostatic potential  $\phi(\boldsymbol{\rho},z)=\varphi(z)e^{i\mathbf{k}\cdot\boldsymbol{\rho}}$ (time dependence $e^{-i\omega t}$ is omitted for clearness).  Expressing the field $\mathbf{E}=-\nabla\phi$ in terms of $\varphi$ gives the in-plane component $\mathbf{E}_{\boldsymbol{\rho}}(\boldsymbol{\rho},z)=-i\mathbf{k}\varphi(z)e^{i\mathbf{k}\cdot\boldsymbol{\rho}}$ and further this field in the ML $\mathbf{E}_{\boldsymbol{\rho}}(\boldsymbol{\rho},0)=-i\mathbf{k}\varphi(0)e^{i\mathbf{k}\cdot\boldsymbol{\rho}}$. Then apply divergence $\nabla_{\boldsymbol{\rho}}$ to  Eq.~(\ref{bigP1}), and we have the polarization charge density $-\nabla\cdot\mathbf{P}=-\delta(z)\nabla_{\boldsymbol{\rho}}\cdot\boldsymbol{\mathcal{P}}=-\delta(z)[a_{21}i\mathbf{k}\cdot\mathbf{w}+a_{22}k^2\varphi(0)e^{i\mathbf{k}\cdot\boldsymbol{\rho}}]$ and obtain  
Poisson's equation,  
\begin{equation}
\nabla^2\phi(\boldsymbol{\rho},z)=4\pi\delta(z)[a_{21}i\mathbf{k}\cdot\mathbf{w}(\boldsymbol{\rho})+a_{22}k^2\varphi(0)e^{i\mathbf{k}\cdot\boldsymbol{\rho}}]. 
\label{poi1}
\end{equation}

To solve Eq.~(\ref{poi1}), we insert the expansions of $\varphi(z)$ [Eq.~(\ref{vphx})] and $\delta(z)$ [Eq.~(\ref{delx})], yielding 
\begin{equation}
\hat{\varphi}(q)=-2\left[a_{21}i\mathbf{k}\cdot\mathbf{w}_0+a_{22}k^2\varphi(0)\right]\frac{1}{k^2+q^2}. 
\label{phiq}
\end{equation}
When this expression is substituted for $\hat{\varphi}(q)$ of the following equation, 
\begin{equation}
\varphi(0)=\int_{-\infty}^{\infty}\hat{\varphi}(q)dq,
\label{vph0x}
\end{equation}
we find  
\begin{equation}
\varphi(0)=-\frac{2\pi a_{21}i\mathbf{k}\cdot\mathbf{w}_0}{k(1+2\pi a_{22}k)},   
\label{vph0a}
\end{equation}
and then obtain the electric field in the ML,  
\begin{equation}
\mathbf{E}_{\boldsymbol{\rho}}(\boldsymbol{\rho},0)=-\frac{2\pi a_{21}\mathbf{k}}{k(1+2\pi a_{22}k)}\mathbf{w}\cdot\mathbf{k}.    
\label{Eroz0}
\end{equation}

For a normal mode with wavevector $\mathbf{k}$,  Eq.~(\ref{eomw1}) becomes 
\begin{equation}
(-\omega^2-a_{11})\mathbf{w}(\boldsymbol{\rho})=a_{12}\mathbf{E}_{\boldsymbol{\rho}}(\boldsymbol{\rho},0).   
\label{eow1b}
\end{equation}

The expression (\ref{Eroz0}) admits two possibilities for $\mathbf{w}\cdot\mathbf{k}$, namely, case (i) $\mathbf{w}\cdot\mathbf{k}=0$, or case (ii) $\mathbf{w}\cdot\mathbf{k}\ne 0$. 
In case (i), as $\mathbf{w}\cdot\mathbf{k}=0$ the normal modes are transverse waves. Eq.~(\ref{vph0a}) gives $\varphi(0)=0$, and on account of Eq.~(\ref{phiq}) we have $\hat{\varphi}(q)=0$. Then it follows from Eq.~(\ref{vphx}) that $\varphi(z)=0$, and therefore in the electrostatic approximation the macroscopic field vanishes identically, $\mathbf{E}(\mathbf{r})=0$. The frequency of the TO mode of wavevector $\mathbf{k}$ is given by the solution of Eq.~(\ref{eow1b}) with the field equal to zero, 
\begin{equation}
\omega_t=\omega_0=\sqrt{-a_{11}}=\sqrt{\frac{K-e_ae_BQ_1}{\bar{m}}},       
\label{wto2a}
\end{equation}
which is independent of wavevector; that is, the long-wavelength TO modes are dispersionless, consistent with previous tight-binding \cite{Sanchez:2002} and first-principles  \cite{Wirtz:2003,Topsakal:2009,Sohier:2017} calculations.    

In case (ii) the electric field $\mathbf{E}_{\boldsymbol{\rho}}(\mathbf{r})$ is nonzero, and from the equations evidently the vectors $\mathbf{w}(\boldsymbol{\rho})$, $\mathbf{E}_{\boldsymbol{\rho}}(\mathbf{r})$, $\mathbf{P}(\mathbf{r})$ associated with  the mode are all longitudinal, i.e., parallel to wave vector $\mathbf{k}$, $\mathbf{w}(\boldsymbol{\rho})\parallel\mathbf{E}_{\boldsymbol{\rho}}(\mathbf{r})\parallel\mathbf{P}(\mathbf{r})\parallel\mathbf{k}$. Inserting expression (\ref{Eroz0}) for $\mathbf{E}_{\boldsymbol{\rho}}(\boldsymbol{\rho},0)$ into Eq.~(\ref{eow1b}) we find the frequency of the longitudinal optical (LO) mode,   
\begin{equation}
\omega_l(k)=\Big(-a_{11}+\frac{2\pi a_{21}^2k}{1+2\pi a_{22}k}\Big)^{1/2}.     
\label{wlo2a}
\end{equation}
Expressing the $a$-coefficients in terms of $e_B$, $\chi_e$, $\omega_0$ by Eqs.~(\ref{a22alf}), (\ref{a21b}) and (\ref{wto2a}), then $\omega_l(k)$ can be rewritten as  
\begin{equation}
\omega_l(k)=\Big[\omega_0^2+\frac{2\pi e_B^2k}{\bar{m}s(1+2\pi\chi_ek)}\Big]^{1/2}.     
\label{wlo2a2}
\end{equation}
Physically, Eq.~(\ref{wlo2a2}) is valid for small wavevectors;  
mathematically $\omega_l$ is a monotonically increasing function of $k$ and has an upper limit $\omega_M$ for very large $k$, 
\begin{equation}
\omega_M=\sqrt{\omega_0^2+e_B^2/(\bar{m}s\chi_e)}~.     
\label{wloupp}
\end{equation}

In the theoretical study by Sohier {\it et al}. \cite{Sohier:2017} the LO phonon dispersion is expressed as $\omega_l^2=\omega_0^2+\mathcal{S}k/(1+r_{eff}k)$.  From their Eqs.~(2) and (3)\footnote{There is a misprint in Eq.~(2) of Ref.\cite{Sohier:2017}: the square should be for each summand rather than for the total sum.} one finds that the parameter $\mathcal{S}$ relates to the Born charge via $\mathcal{S}=2\pi e_B^2/(\bar{m}s)$. $r_{eff}$ is an effective screening length, given by by $r_{eff}=\epsilon_pt/2$ with an effective medium model \cite{Sohier:2016,Sohier:2017}, where $\epsilon_p$ and $t$ are effective dielectric constant and effective thickness of the ML material. Both parameters $\mathcal{S}$ and $r_{eff}$ are computed by first-principles calculation \cite{Sohier:2017}.  According to the defintion in Refs.\cite{Cudazzo:2011,Berkelbach:2013}, the effective screening length is given by $2\pi\chi_e$ or $2\pi\chi_0$ when including the vibrational contribution [$\chi_0$ is the static susceptibility  Eq.~(\ref{chiwps0})]. As it is the {\it high-frequency} dielectric constant that determines the difference between the squared LO and TO phonon frequencies \cite{Cochran:1962,Giannozzi:1991,Gonze:1997} so  $r_{eff}$ takes the former, i.e., $r_{eff}=2\pi\chi_e$. Therefore the LO phonon dispersion Eq.~(\ref{wlo2a2}) is identical to the analytical expression of Sohier {\it et al}..

Having $\varphi(0)$ [Eq.~(\ref{vph0a})] for the LO mode $\omega_l(k)$, we substitute expression (\ref{phiq}) back into Eq.~(\ref{vphx}) to give the electric potential and then obtain the macroscopic field associated with the LO mode, expressed in terms of mode displacement $\mathbf{w}$ [in a form consistent with  Eqs.~(\ref{Ero1}) and (\ref{Ez1})], 
\begin{subequations} 
\begin{equation}
\mathbf{E}_{\boldsymbol{\rho}}(\boldsymbol{\rho},z)=-\frac{2\pi e_B\mathbf{k}}{\sqrt{\bar{m}s}k(1+2\pi \chi_ek)}\mathbf{w}\cdot\mathbf{k}e^{-k\lvert z\rvert}, 
\label{Erolo}
\end{equation}
\begin{equation}
\mathbf{E}_z(\boldsymbol{\rho},z)=-\mathbf{e}_z\frac{2\pi ie_B}{\sqrt{\bar{m}s}(1+2\pi \chi_ek)}\mathbf{w}\cdot\mathbf{k}\sgn(z)e^{-k\lvert z\rvert}.  
\label{Ezlo}
\end{equation}
\end{subequations}

The long-range electrostatic interactions cause a higher LO frequency than the TO frequency for a finite $k$ \cite{Sanchez:2002} with the splitting determined by the latter term in the square brackets of Eq.~(\ref{wlo2a2}). This term also shows that there is no splitting in the limit $k\rightarrow 0$ as the macroscopic field vanishes [Eq.~(\ref{Erolo})], which is different from the situation in bulk BN where LO-TO splitting occurs also at the $\Gamma$ point \cite{Ohba:2001,Topsakal:2009}.  Therefore, the transparent expressions (\ref{wto2a}) and (\ref{wlo2a2}) show the degeneracy of the LO and TO modes at $\Gamma$ and their splitting at a finite wavevector, well-known  phenomena of the 2D semiconductors \cite{Sanchez:2002,Wirtz:2003,Michel:2009,Topsakal:2009,Sohier:2017}.

The out-of-plane optical vibrations, namely, the ZO modes, can be obtained from Eqs.~(\ref{eomw1z}) and (\ref{bigP1z}) with the electrostatic approach as follows. Recall that the $z$-component of the field [Eq.~(\ref{Ez2})] has a $\delta(z)$ term, divergent at $z=0$, because the ML is treated as a geometric plane where the ionic charge distribution and polarization density $\mathbf{P}$ have a $\delta(z)$ form. We approximate $\delta(z)$ by a Gaussian distribution with a small thickness $\varepsilon$ 
($\varepsilon\rightarrow 0$ ), $\delta_{\varepsilon}(z)=\frac{1}{\sqrt{\pi}\varepsilon}e^{-z^2/\varepsilon^2}$, as in the theoretical study \cite{Michel:2009}. A  similar treatment is used also in first-principles calculations \cite{Sohier:2017}. 
The $z$-component of the field in the ML is $\mathbf{E}_z(\boldsymbol{\rho},0)=-\mathbf{e}_z\varphi'(0)e^{i\mathbf{k}\cdot\boldsymbol{\rho}}$, and after inserting this field into Eq.~(\ref{bigP1z}) the polarization charge density can be expressed as $-\nabla\cdot\mathbf{P}=-\delta_{\varepsilon}'(z)\mathbf{e}_z\cdot\boldsymbol{\mathcal{P}}=-[c_{21}\mathbf{w}\cdot\mathbf{e}_z-c_{22}\varphi'(0)e^{i\mathbf{k}\cdot\boldsymbol{\rho}}]\delta_{\varepsilon}'(z)$.  Thus Poisson's equation is given by  
\begin{equation}
\nabla^2\phi(\boldsymbol{\rho},z)=4\pi\delta_{\varepsilon}'(z)[c_{21}\mathbf{w}(\boldsymbol{\rho})\cdot\mathbf{e}_z-c_{22}\varphi'(0)e^{i\mathbf{k}\cdot\boldsymbol{\rho}}]. 
\label{poiz1}
\end{equation}

Expanding $\varphi(z)$ [Eq.~(\ref{vphx})] and $\delta_{\varepsilon}(z)$ 
\begin{equation}
\delta_{\varepsilon}(z)=\frac{1}{2\pi}\int_{-\infty}^{\infty}e^{-(q\varepsilon/2)^2}e^{iqz}dq,  
\label{deldx}
\end{equation}
then we have 
\begin{equation}
\varphi'(z)=\int_{-\infty}^{\infty}iq\hat{\varphi}(q)e^{iqz}dq,
\label{vph1x}
\end{equation}
\begin{equation}
\delta_{\varepsilon}'(z)=\frac{1}{2\pi}\int_{-\infty}^{\infty}iqe^{-(q\varepsilon/2)^2}e^{iqz}dq. 
\label{deld1x}
\end{equation}
Inserting the expansions (\ref{vphx}) and (\ref{deld1x}) into Poisson's equation (\ref{poiz1}) we find $\hat{\varphi}(q)$,  
\begin{equation}
\hat{\varphi}(q)=-2\left[c_{21}\mathbf{w}_0\cdot\mathbf{e}_z-c_{22}\varphi'(0)\right]\frac{iq}{k^2+q^2}e^{-(q\varepsilon/2)^2},  
\label{phiqz}
\end{equation}
and then carrying out the integration in Eq.~(\ref{vph1x}) we obtain 
\begin{align}
\varphi'(z) &=2\Big[c_{21}\mathbf{w}_0\cdot\mathbf{e}_z-c_{22}\varphi'(0)\Big]\left\{\frac{2\sqrt{\pi}}{\varepsilon}e^{-z^2/\varepsilon^2}\right.  
\nonumber \\ &\qquad {}  -\frac{\pi}{2}ke^{(k\varepsilon/2)^2}\Big[2\cosh(kz)-e^{-k\lvert z\rvert}\erf\big(\frac{k\varepsilon}{2}-\frac{\lvert z\rvert}{\varepsilon}\big)
\Big. 
\nonumber \\ &\qquad {} \left. \Big. 
-e^{k\lvert z\rvert}\erf\big(\frac{k\varepsilon}{2}+\frac{\lvert z\rvert}{\varepsilon}\big)\Big]  \right\},  
\label{phi1za}
\end{align}
where $\erf(x)$ is the error function, $\erf(x)=\frac{2}{\sqrt{\pi}}\int_0^xe^{-t^2}dt$. 

Taking $z=0$ in Eq.~(\ref{phi1za}) we find $\varphi'(0)$, 
\begin{equation}
\varphi'(0)=\frac{c_{21}}{\varepsilon/\zeta_k+c_{22}}\mathbf{w}_0\cdot\mathbf{e}_z,   
\label{phi1z0b}
\end{equation}
where 
\begin{equation}
\zeta_k=4\sqrt{\pi}\Big[1-\sqrt{\pi}\frac{k\varepsilon}{2}\erfc(\frac{k\varepsilon}{2})e^{(k\varepsilon/2)^2}\Big],   
\label{zetak}
\end{equation}
$\erfc(x)$ being the complementary error function, $\erfc(x)=1-\erf(x)$. 

The electric field in the ML $\mathbf{E}_z(\boldsymbol{\rho},0)$ follows,   
\begin{equation}
\mathbf{E}_z(\boldsymbol{\rho},0)=-\frac{c_{21}}{\varepsilon/\zeta_k+c_{22}}\mathbf{w}=-\frac{e_B^\prime}{\sqrt{\bar{m}s}(\varepsilon/\zeta_k+\chi_e^\prime)}\mathbf{w}. 
\label{Ezro0b1}
\end{equation}
When this field is substituted into the equation of motion (\ref{eomw1z}), we obtain the frequency of the ZO mode,
\begin{equation}
\omega_z(k)=\Big[\omega_0'^2+\frac{e_B^{\prime 2}}{\bar{m}s(\varepsilon/\zeta_k+\chi_e^\prime)}\Big]^{1/2}.     
\label{wzo2a}
\end{equation}

Considering that both $\varepsilon$ and $k$ are small quantities, expression (\ref{zetak}) reduces to a constant $\zeta=4\sqrt{\pi}$, 
and the ZO phonon frequency becomes independent of wavevector,   
\begin{equation}
\omega_z=\Big\{\omega_0'^2+\frac{e_B^{\prime 2}}{\bar{m}s[\varepsilon/(4\sqrt{\pi})+\chi_e^\prime]}\Big\}^{1/2};       
\label{wzo2a2}
\end{equation}
for a planar material when taking $\varepsilon \ll 4\sqrt{\pi}\chi_e^\prime$ the frequency becomes  
\begin{equation}
\omega_z=\Big(\omega_0'^2+\frac{e_B^{\prime 2}}{\bar{m}s\chi_e^\prime}\Big)^{1/2}.     
\label{wzo2b3}
\end{equation}
The ionic polarization (i.e., $\chi_e^\prime\ne 0$) ensures a finite frequency;  otherwise $\omega_z$ [Eq.~(\ref{wzo2a2})] becomes extremely large and even 
$\omega_z\rightarrow \infty$ [Eq.~(\ref{wzo2b3})] when $\chi_e^\prime$ is neglected in the RIM (also see Table~\ref{table:4}).  
The result here that the long wavelength ZO modes are dispersionless agrees with previous calculations which show a very flat dispersion curve at small wavevectors  \cite{Sanchez:2002,Wirtz:2003,Michel:2009,Topsakal:2009}. 

Having $\varphi'(0)$ [Eq.~(\ref{phi1z0b})] for a ZO mode we then get $\varphi'(z)$ from  Eq.~(\ref{phi1za}) and also obtain $\varphi(z)$ after substituting the $\hat{\varphi}(q)$ expression (\ref{phiqz}) into Eq.~(\ref{vphx}). It follows that the macroscopic field associated with the ZO mode is given by 
\begin{subequations} 
\begin{align}
\mathbf{E}_{\boldsymbol{\rho}}(\boldsymbol{\rho},z)&=\frac{i\pi e_B^\prime\mathbf{k}}{\sqrt{\bar{m}s}(1+\chi_e^\prime\zeta_k/\varepsilon)}\mathbf{w}\cdot\mathbf{e}_z\sgn(z)e^{(k\varepsilon/2)^2}  \nonumber \\ 
&\qquad {}  \times\Big[2\sinh(k\lvert z\rvert)+e^{-k\lvert z\rvert}\erf\big(\frac{k\varepsilon}{2}-\frac{\lvert z\rvert}{\varepsilon}\big)  \Big. 
\nonumber \\ &\qquad {} \Big. -e^{k\lvert z\rvert}\erf\big(\frac{k\varepsilon}{2}+\frac{\lvert z\rvert}{\varepsilon}\big)\Big],  
\label{Erozo8}
\end{align}
\begin{align}
\mathbf{E}_z(\boldsymbol{\rho},z)&=\frac{-2e_B^\prime}{\sqrt{\bar{m}s}(1+\chi_e^\prime\zeta_k/\varepsilon)}\mathbf{w}\left\{\frac{2\sqrt{\pi}}{\varepsilon}e^{-z^2/\varepsilon^2}  \right. \nonumber \\ 
&\qquad {}   
-\frac{\pi}{2}ke^{(k\varepsilon/2)^2}\Big[2\cosh(kz)-e^{-k\lvert z\rvert}\erf\big(\frac{k\varepsilon}{2}-\frac{\lvert z\rvert}{\varepsilon}\big) \Big. 
\nonumber \\  &\qquad {}  \left. \Big. -e^{k\lvert z\rvert}\erf\big(\frac{k\varepsilon}{2}+\frac{\lvert z\rvert}{\varepsilon}\big)\Big]  \right\},     
\label{Ezzo8}
\end{align}
\end{subequations} 
which for small $k$ and $\varepsilon$ reduce to the following expressions [compare to Eqs.~(\ref{Ez2}) and (\ref{Ero2})], 
\begin{subequations} 
\begin{equation}
\mathbf{E}_{\boldsymbol{\rho}}(\boldsymbol{\rho},z)=\frac{-2i\pi e_B^\prime\mathbf{k}}{\sqrt{\bar{m}s}(1+\chi_e^\prime 4\sqrt{\pi}/\varepsilon)}\mathbf{w}\cdot\mathbf{e}_z\sgn(z)e^{-k\lvert z\rvert},  
\label{Erozo9}
\end{equation}
\begin{equation}
\mathbf{E}_z(\boldsymbol{\rho},z)=\frac{-4\pi e_B^\prime}{\sqrt{\bar{m}s}(1+\chi_e^\prime 4\sqrt{\pi}/\varepsilon)}\mathbf{w}\left(\delta(z)-\frac{1}{2}ke^{-k\lvert z\rvert}\right).    
\label{Ezzo9}
\end{equation}
\end{subequations} 
In particular we note that the in-plane component is an odd function of $z$, and  negligible at very small $k$ and $\varepsilon$.

\subsection{Phonon group velocity and density of states}

Expanding $\omega_l(k)$ [Eq.~(\ref{wlo2a2})] to second order in wavevector $k$,  the LO phonon frequency near the $\Gamma$ point is given by 
\begin{equation}
\omega_l(k)=\omega_0+c_lk-\frac{1}{2}c_l(\frac{c_l}{\omega_0}+4\pi \chi_e)k^2,     
\label{wlok0}
\end{equation}
where  
\begin{equation}
c_l=\frac{\pi e_B^2}{\bar{m}s\omega_0},     
\label{cl}
\end{equation}
which is identical to that given in Ref.\cite{Michel:2009}. 

From the dispersion Eq.~(\ref{wlo2a2}) we find straightforward the DOS of the LO modes,  
\begin{equation}
g_l(\omega)=\frac{c_l\omega_0}{4\pi^4 \chi_e^3}\frac{\omega(\omega^2-\omega_0^2)}{\big(\omega_M^2-\omega^2\big)^3}, \quad \omega_0 \leq \omega < \omega_M,   
\label{doslo1}
\end{equation}
where $\omega_M$ is the upper bound of LO phonon frequency [expression~(\ref{wloupp})].  

Similarly, expanding $\omega_z(k)$ [Eq.~(\ref{wzo2a})] we find the ZO phonon frequency near $\Gamma$, 
\begin{equation}
\omega_z(k)=\omega_z-c_zk-\frac{1}{2}c_z\Big(\frac{c_z}{\omega_z}+ \frac{4\pi\varepsilon\chi_e^\prime}{\varepsilon+4\sqrt{\pi}\chi_e^\prime}-\frac{2\varepsilon}{\sqrt{\pi}}\Big)k^2,     
\label{wzok0}
\end{equation}
where $\omega_z$ is the ZO mode frequency given by Eq.~(\ref{wzo2a2}), 
and 
\begin{equation}
c_z=\frac{\pi e_B'^2}{\bar{m}s\omega_z}\Big(\frac{\varepsilon}{\varepsilon+4\sqrt{\pi}\chi_e^\prime}\Big)^2.      
\label{cz}
\end{equation}
Dropping the terms due to the EP of ions, expansions (\ref{wlok0}) and (\ref{wzok0}) then reduce to a form  as given by the RIM of Ref.\cite{Michel:2009} [Eqs.~(57) and (59) therein]. 
The nearly flat dispersion of the ZO modes allows their DOS to be approximated by \cite{Michel:2009} 
\begin{equation}
g_z(\omega)=\frac{\omega_z-\omega}{2\pi c_z^2},  \quad   \omega < \omega_z.   
\label{doszo1}
\end{equation}
Evidently $c_z$ is close to zero as $\varepsilon \ll 4\sqrt{\pi}\chi_e^\prime$ for the 2D crystal, resulting in the ZO modes being nondispersive. 

The group velocity is $\nabla_{\mathbf{k}}\omega_i(k)=\omega_i'(k)\mathbf{k}/k$, where $i$ indexes a phonon branch of LO, TO or ZO modes so the $c_l$ and $c_z$ above are simply the norms of the group velocities of the LO and ZO modes at $\Gamma$,  respectively, corresponding to the slopes \cite{Sohier:2017} of the phonon dispersion curves. For the long lattice waves, as the wavevector $k$ increases, clearly $\omega_l$ increases and $\omega_z$ decreases, whereas  
$\omega_t$ stays flat near $\Gamma$ [Eq.~(\ref{wto2a})].  

\subsection{2D lattice dielectric susceptibility and dielectric function}

The lattice susceptibility of ML hBN for in-plane polarization can be deduced from Eqs.~(\ref{eomw1}) and (\ref{bigP1}) by considering  periodic solutions, $\mathbf{W}$, $\mathbf{E}$, $\boldsymbol{\mathcal{P}}$ $\propto e^{-i\omega t}$, due to an external charge or electromagnetic field with oscillation frequency $\omega$.  Then Eq.~(\ref{eomw1})  reduces simply to Eq.~(\ref{eow1b}). Inserting Eq.~(\ref{eow1b}) into Eq.~(\ref{bigP1}) to eliminate $\mathbf{W}$, one finds 
\begin{equation}
\boldsymbol{\mathcal{P}}(\boldsymbol{\rho})=[a_{22}-\frac{a_{12}a_{21}}{\omega^2+a_{11}}]\mathbf{E}_{\boldsymbol{\rho}}(\boldsymbol{\rho},0),      
\label{bgPfroE}
\end{equation}
and subsequently the 2D dielectric susceptibility, defined by $\chi=\boldsymbol{\mathcal{P}}(\boldsymbol{\rho})/\mathbf{E}_{\boldsymbol{\rho}}(\boldsymbol{\rho},0)$, 
\begin{equation}
\chi(\omega)=a_{22}-\frac{a_{12}a_{21}}{\omega^2+a_{11}}=\chi_e+\frac{e_B^2}{\bar{m}s(\omega_0^2-\omega^2)},          
\label{chiwp1}
\end{equation}
where $\chi_e$ is the high-frequency (i.e., clamped-ion, or electronic) susceptibility.  

The static dielectric susceptibility $\chi_0$ follows from expression (\ref{chiwp1}),  
\begin{equation}
\chi_0=\chi_e+\frac{e_B^2}{\bar{m}s\omega_0^2},       
\label{chiwps0}
\end{equation}
where the second term on the right-hand side is the vibrational (also called ionic) contribution. 
With Eq.~(\ref{chiwps0}), $a_{12}$ [Eq.~(\ref{a12})] can be expressed in terms of the 2D susceptibilities, 
\begin{equation}
a_{12}=a_{21}=\omega_0\sqrt{\chi_0-\chi_e}.        
\label{a12chi09}
\end{equation}

The 2D dynamical susceptibility $\chi(\omega)$ can be expressed in terms of three quantities that are usually used, namely,  
the intrinsic oscillator frequency $\omega_0$ and the high-frequency and static susceptibilities $\chi_e$, $\chi_0$, 
\begin{equation}
\chi(\omega)=\chi_e+\frac{\chi_0-\chi_e}{1-\omega^2/\omega_0^2}~,       
\label{chiwp2}
\end{equation}
which has a similar form to its counterpart of bulk polar crystals \cite{Born:1954}.   

Similarly, the 2D dielectric susceptibility for the vertical $z$-polarization $\chi'$ (the prime does not indicate a derivative), defined by $\chi'=\boldsymbol{\mathcal{P}}(\boldsymbol{\rho})/\mathbf{E}_z(\boldsymbol{\rho},0)$ for any particular $\omega$, is deduced from Eqs.~(\ref{eomw1z}) and (\ref{bigP1z}), 
\begin{equation}
\chi'(\omega)=c_{22}-\frac{c_{12}c_{21}}{\omega^2+c_{11}}=\chi_e'+\frac{e_B'^2}{\bar{m}s(\omega_0'^2-\omega^2)},          
\label{chiwz1}
\end{equation}
with the static dielectric susceptibility $\chi_0'$,    
\begin{equation}
\chi_0'=\chi_e'+\frac{e_B'^2}{\bar{m}s\omega_0'^2}.         
\label{chiwzs0}
\end{equation}

Combining Eqs.~(\ref{chiwzs0}) and (\ref{c12z}), one can express $c_{12}$ in terms of the 2D susceptibilities, 
\begin{equation}
c_{12}=c_{21}=\omega_0'\sqrt{\chi_0'-\chi_e'}.        
\label{c12chi09}
\end{equation}

The 2D dynamical susceptibility $\chi'(\omega)$ can also be expressed in terms of three quantities $\chi_e'$, $\chi_0'$, $\omega_0'$, 
\begin{equation}
\chi'(\omega)=\chi_e'+\frac{\chi_0'-\chi_e'}{1-\omega^2/\omega_0'^2}.      
\label{chiwz2}
\end{equation}

On eliminating $e_B'^2/(\bar{m}s)$ in Eq.~(\ref{wzo2b3}) with the help of Eq.~(\ref{chiwzs0}) we find a simple relation 
\begin{equation}
\frac{\omega_z^2}{\omega_0'^2}=\frac{\chi_0'}{\chi_e'}.      
\label{wz02chz0}
\end{equation}

The 2D high-frequency and static dielectric susceptibilities can be calculated from first principles, and knowing their values the susceptibility expressions (\ref{chiwps0}) and (\ref{chiwzs0}) obtained here will be used below in Sec. III to determine the model parameters such as the effective charges and spring force constants, and the $a$- or $c$-coefficients of the lattice equations 
[Eqs.~(\ref{eomw1co}), (\ref{bigP1co}), (\ref{eomw1zco}) and (\ref{bigP1zco})].

The lattice dielectric function (DF) can be obtained by considering the response of the 2D lattice to a test  charge in the electrostatic approximation. Now the equation of electrostatics is given by $\nabla\cdot(\mathbf{E}+4\pi\mathbf{P})=4\pi\sigma$, where $\sigma$ is the test charge density function dependent on $\mathbf{r}$, $t$, and $\mathbf{E}$ is the total field of the test charge plus the polarization charge. To find the dielectric response let $\sigma$ have the form $\sigma=\sigma_0e^{i\mathbf{k}\cdot\boldsymbol{\rho}}f(z)$, where $f(z)$ is an arbitrary function to describe the charge distribution in the $z$ direction, for the general case that the external charge is not necessarily confined in the ML, and time dependence $e^{-i\omega t}$ has been absorbed into $\sigma_0$ for clearness.

First, let us consider a charge distribution that is asymmetric with respect to the ML, i.e., $f(-z)\neq f(z)$, for instance, when the test charge is simply put above or below the ML. Then the field due to the charge is nonzero in the ML, $\mathbf{E}_{\boldsymbol{\rho}}(\boldsymbol{\rho},0)\neq 0$, $\mathbf{E}_z(\boldsymbol{\rho},0)\neq 0$, and therefore the lattice responds creating both in-plane [Eqs.~(\ref{eomw1co}) and (\ref{bigP1co})] and out-of-plane [Eqs.~(\ref{eomw1zco}) and (\ref{bigP1zco})] vibrations, with all quantities such as $\mathbf{w}_{\boldsymbol{\rho}}$, $\mathbf{w}_z$,  $\boldsymbol{\mathcal{P}}_{\boldsymbol{\rho}}$, $\boldsymbol{\mathcal{P}}_z$, $\mathbf{E}$ varying according to $e^{i(\mathbf{k}\cdot\boldsymbol{\rho}-\omega t)}$. 
Now the dielectric polarization is given by the sum of the in-plane and out-of-plane contributions, $\mathbf{P}=(\boldsymbol{\mathcal{P}}_{\boldsymbol{\rho}}+\boldsymbol{\mathcal{P}}_z)\delta(z)$, where $\boldsymbol{\mathcal{P}}_{\boldsymbol{\rho}}=\chi(\omega)\mathbf{E}_{\boldsymbol{\rho}}(\boldsymbol{\rho},0)$ and $\boldsymbol{\mathcal{P}}_z=\chi'(\omega)\mathbf{E}_z(\boldsymbol{\rho},0)$, with $\chi$ and $\chi'$ being the 2D susceptibilities [Eqs.~(\ref{chiwp1}) and (\ref{chiwz1})]. Writing $\mathbf{E}=-\nabla\phi$ with the total electrostatic potential $\phi(\boldsymbol{\rho},z)=\varphi(z)e^{i\mathbf{k}\cdot\boldsymbol{\rho}}$, one expresses the polarization charge density $-\nabla\cdot\mathbf{P}$ in terms of the  derivatives of potential $\phi$, and then Poisson's equation follows from the equation of electrostatics, 
\begin{align}
\nabla^2\phi(\boldsymbol{\rho},z)&=-4\pi\left[\sigma_0 e^{i\mathbf{k}\cdot\boldsymbol{\rho}}f(z)+\chi(\omega)\delta(z)
\nabla_{\boldsymbol{\rho}}^2\phi(\boldsymbol{\rho},0) \right.  
\nonumber \\ &\qquad {}
\left. +\chi'(\omega)\varphi'(0)e^{i\mathbf{k}\cdot\boldsymbol{\rho}}\delta'(z)\right], 
\label{poions1}
\end{align}
where $\phi(\boldsymbol{\rho},0)$ is the total potential in the ML, $\phi(\boldsymbol{\rho},0)=\varphi(0)e^{i\mathbf{k}\cdot\boldsymbol{\rho}}$.  
Inserting the expansion of $f(z)$, 
\begin{equation}
f(z)=\int_{-\infty}^{\infty}f(q)e^{iqz}dq,
\label{fzex}
\end{equation} 
and expansions (\ref{vphx}) and (\ref{delx}) into Poisson's equation (\ref{poions1}) one finds $\hat{\varphi}(q)$, 
\begin{equation}
\hat{\varphi}(q)=\frac{2}{k^2+q^2}\left[ 2\pi\sigma_0f(q)-\chi(\omega)k^2\varphi(0)+i\chi'(\omega)\varphi'(0)q \right],       
\label{vaphqdcp}
\end{equation}
where the three terms on the right-hand side represent the respective contributions due to the external charge, in-plane and out-of-plane lattice polarization. Here $\hat{\varphi}(q)$ is expressed in terms of $\varphi(0)$ and $\varphi'(0)$ as they are proportional to the field components in the ML, i.e., 
$\varphi(0)\propto \mathbf{E}_{\boldsymbol{\rho}}(\boldsymbol{\rho},0)$ and $\varphi'(0)\propto \mathbf{E}_z(\boldsymbol{\rho},0)$. 
 Integrating $\hat{\varphi}(q)$ over $q$ according to Eq.~(\ref{vph0x}), one then obtains $\varphi(0)$,  
\begin{equation}
\varphi(0)=\frac{\varphi_\sigma(0)}{1+2\pi k\chi(\omega)}~,      
\label{vaphz0dc}
\end{equation}
where 
\begin{equation}
\varphi_\sigma(0)=4\pi\sigma_0\int_{-\infty}^{\infty}\frac{f(q)}{k^2+q^2}dq~.       
\label{vafexz0}
\end{equation}

The out-of-plane (ZO) motion makes no contribution to $\varphi(0)$ as the integration value of the third term of $\hat{\varphi}(q)$ is zero. 
Clearly $\varphi_\sigma(0)e^{i\mathbf{k}\cdot\boldsymbol{\rho}}$ is the electrostatic potential in the ML due to the test charge alone, and therefore the  dielectric function of the 2D lattice, which is defined as the ratio of this test charge potential to the total potential in the ML $\varphi(0)e^{i\mathbf{k}\cdot\boldsymbol{\rho}}$ (extension of the 2D {\it wavevector}-dependent dielectric function $\epsilon(k)$ of Ref.\cite{Cudazzo:2011}), is given by 
\begin{equation}
\epsilon(k,\omega)=1+2\pi k\chi(\omega)=1+2\pi k\left[\chi_e+\frac{e_B^2}{\bar{m}s(\omega_0^2-\omega^2)}\right],     
\label{df2D}
\end{equation}
showing a linear dependence on wavevector as the $\epsilon(k)$ of Ref.\cite{Cudazzo:2011}).

In expression~(\ref{vaphqdcp}) for $\hat{\varphi}(q)$,  
 $\varphi(0)$ is known [Eq.~(\ref{vaphz0dc})], and $\varphi'(0)$ can be determined 
through expression~(\ref{vph1x}) for $\varphi'(z)$ as follows. Substituting expression~(\ref{vaphqdcp}) for $\hat{\varphi}(q)$ in Eq.~(\ref{vph1x}) one finds
\begin{align}
\varphi'(z) &=4\pi\sigma_0\int_{-\infty}^{\infty}\frac{iqf(q)}{k^2+q^2}e^{iqz}dq \nonumber \\
&\qquad {} -2k^2\chi(\omega)\varphi(0)\int_{-\infty}^{\infty}\frac{iq}{k^2+q^2}e^{iqz}dq \nonumber \\ 
&\qquad {} -4\pi\chi'(\omega)\varphi'(0)\Big[\delta(z)-\frac{1}{2}ke^{-k\lvert z\rvert}\Big]. 
\label{vaph1zdfa}
\end{align}
The square bracketed part is a familiar form that has appeared in Eq.~(\ref{Ez2}). To find $\varphi'(0)$ we introduce an effective thickness $\varepsilon$ to $\delta(z)$ to  approach it with $\delta_{\varepsilon}(z)$ as was done above in Sec. III.  Let $\Lambda=\delta_{\varepsilon}(0)$; $\Lambda=1/(\sqrt{\pi}\varepsilon)$, for instance, when $\delta(z)$ is approximated by a Gaussian \cite{Michel:2009}, $\delta_{\varepsilon}(z)=e^{-z^2/\varepsilon^2}/(\sqrt{\pi}\varepsilon)$.  
Now taking $z=0$ in Eq.~(\ref{vaph1zdfa}), the term containing $\chi(\omega)\varphi(0)$ vanishes, i.e., the in-plane motion makes no contribution to 
$\varphi'(0)$, which is given by 
\begin{equation}
\varphi'(0)=\frac{4\pi\sigma_0}{1+4\pi\chi'(\omega)(\Lambda-k/2)}\int_{-\infty}^{\infty}\frac{iqf(q)}{k^2+q^2}dq~. 
\label{vaph1z0dfb}
\end{equation}

Having $\hat{\varphi}(q)$ [expression~(\ref{vaphqdcp})] then the total potential $\phi$ is known, and the total field $\mathbf{E}$ can be obtained straightforward, 
\begin{subequations} 
\begin{align}
\mathbf{E}_{\boldsymbol{\rho}}(\boldsymbol{\rho},z)&=-2\pi i\mathbf{k}\Big\{2\sigma_0\int_{-\infty}^{\infty}\frac{f(q)e^{iqz}}{k^2+q^2}dq -[ k\chi(\omega)\varphi(0) \Big. 
\nonumber \\ &\qquad {} \Big.+\chi'(\omega)\varphi'(0)\sgn(z)]e^{-k\lvert z\rvert}\Big\}e^{i\mathbf{k}\cdot\boldsymbol{\rho}}, 
\label{fiedxydfc1}
\end{align}
\begin{align}
\mathbf{E}_z(\boldsymbol{\rho},z) &=-2\pi\mathbf{e}_z\left\{2\sigma_0\int_{-\infty}^{\infty}\frac{iqf(q)}{k^2+q^2}e^{iqz}dq  \right. 
\nonumber \\ &\qquad {} +k^2\chi(\omega)\varphi(0)\sgn(z)e^{-k\lvert z\rvert} 
\nonumber \\ 
&\qquad {} \left. -2\chi'(\omega)\varphi'(0)\Big[\delta(z)-\frac{1}{2}ke^{-k\lvert z\rvert}\Big]\right\}e^{i\mathbf{k}\cdot\boldsymbol{\rho}}~,  
\label{fiedzdfc1}
\end{align}
\end{subequations} 
where $\varphi(0)$ and $\varphi'(0)$ are given by expressions~(\ref{vaphz0dc}) and  (\ref{vaph1z0dfb}), respectively.

 Recalling $\mathbf{E}_{\boldsymbol{\rho}}(\boldsymbol{\rho},z)=-i\mathbf{k}\phi(\mathbf{r})$ with Eq.~(\ref{fiedxydfc1}) for $\mathbf{E}_{\boldsymbol{\rho}}(\boldsymbol{\rho},z)$, evidently the induced potential associated with the ZO polarization is proportional to  $\chi'(\omega)\varphi'(0)\sgn(z)e^{-k\lvert z\rvert}$, which is zero in the ML plane and {\it antisymmetric} with respect to it, consistent with Eq.~(\ref{Pophez}) above. 
The in-plane component of the field associated with the ZO motion [the term containing $\chi'(\omega)\varphi'(0)$ in Eq.~(\ref{fiedxydfc1})] is zero in the ML and thus the ZO motion does not influence the LO motion; meanwhile the LO motion has zero $z$-component of its field [the $\chi(\omega)\varphi(0)$ term in Eq.~(\ref{fiedzdfc1})] in the ML and thus no influence on the ZO motion, making the LO and ZO vibrations driven by the exernal charge essentially independent of each other.  
For small thickness $\varepsilon$, $\varphi'(0)$ is small from Eq.~(\ref{vaph1z0dfb})  
and there is only a small ZO component, which becomes negligible in the limit 
$\varepsilon\rightarrow 0$, in the driven lattice motion [refer to Eq.~(\ref{poions1})]. 
For a symmetric test charge distribution $f(-z)=f(z)$, $\mathbf{E}_z(\boldsymbol{\rho},0)=0$, and this causes only in-plane vibration, with displacement $\mathbf{w}$ and polarization $\mathbf{P}$ both parallel to the ML plane, and evidently yields the same DF $\epsilon(k,\omega)$ as given by expression (\ref{df2D}) above.

As the polarization charge density 
associated with the in-plane motion is $-\delta(z)\nabla_{\boldsymbol{\rho}}\cdot\boldsymbol{\mathcal{P}}_{\boldsymbol{\rho}}(\boldsymbol{\rho})=-\delta(z)[a_{21}-a_{22}(\omega^2+a_{11})/a_{12}]i\mathbf{k}\cdot\mathbf{w}_{\boldsymbol{\rho}}\ne 0$, and $\mathbf{w}_{\boldsymbol{\rho}}\parallel\boldsymbol{\mathcal{P}}_{\boldsymbol{\rho}}\parallel\mathbf{E}_{\boldsymbol{\rho}}(\boldsymbol{\rho},0)\parallel\mathbf{k}$, the DF $\epsilon(k,\omega)$ is due only to the LO vibrations and thus is a longitudinal DF. In the absence of a test charge ($\sigma=0$), there is still a finite electric field $\mathbf{E}_{\boldsymbol{\rho}}(\boldsymbol{\rho},0)$ and potential $\phi(\boldsymbol{\rho},0)$ due to the LO vibrations, and therefore it follows from Eq.~(\ref{vaphz0dc}) that the LO modes are the solutions to 
\begin{equation}
\epsilon(k,\omega)=0.       
\label{dfwl0}
\end{equation}

From Eqs.~(\ref{df2D}) and (\ref{dfwl0}) one indeed obtains the LO phonon dispersion $\omega_l(k)$ of expression (\ref{wlo2a2}). In bulk crystals the lattice DF is zero at the bulk LO mode frequency \cite{Born:1954}.  
From expression (\ref{df2D}) the static DF is $\epsilon_0(k)=1+2\pi\chi_0k$,   
while at high frequencies $\omega\gg\omega_0$ the vibrational contribution is negligible yielding the DF  
 $\epsilon_{\infty}(k)=1+2\pi\chi_ek$, with $2\pi\chi_0$ and $2\pi\chi_e$ being the   effective screening lengths \cite{Cudazzo:2011,Berkelbach:2013}. Both static and high-frequency DFs have the same form as that deduced by Cudazzo {\it et al}. \cite{Cudazzo:2011}.

In terms of the three key quantities $\omega_t$, $\omega_l(k)$, $\epsilon_{\infty}(k)$ the lattice DF of the 2D crystal can be expressed in the form 
\begin{equation}
\epsilon(k,\omega)=\epsilon_{\infty}(k)\frac{\omega^2-\omega_l^2(k)}{\omega^2-\omega_t^2},       
\label{df2D2}
\end{equation}
which is similar to the lattice DF of bulk polar crystals \cite{Born:1954}, the difference being that here both the LO phonon frequency and DF are dependent on   wavevector. 
 In bulk crystals there is the LST relation  \cite{Lyddane:1941}, $\omega_l^2/\omega_t^2=\epsilon_0/\epsilon_{\infty}$;  
for the 2D crystal a similar relation can be deduced from expression (\ref{df2D2}), 
\begin{equation}
\frac{\omega_l^2(k)}{\omega_t^2}=\frac{\epsilon_0(k)}{\epsilon_{\infty}(k)},       
\label{LST2}
\end{equation}
with the three quantities dependent on wavevector.

The extended LST relation (\ref{LST2}) connects the phonon frequencies to the two DFs; given the former, then the ratio of the latter is known, and vice versa.  Furthermore, since the difference between the LO and TO frequencies is due entirely to the macroscopic field, the relation can be used  to measure the phonon frequency change caused by the field. 
Similarly, for out-of-plane motion the susceptibilities $\chi_0'$ and $\chi_e'$ are related to phonon frequencies $\omega_z$ and $\omega_0'$ via the frequency--susceptibility relation (\ref{wz02chz0}), which conveniently quantifies the effect of the macroscopic field on the ZO phonon frequency.

\section{Lattice dynamical properties: local field and polarizable ion effects}

\subsection{In-plane optical modes}

The 2D clamped-ion susceptibility $\chi_e$, a key quantity entering the derived expressions for in-plane motion, can be obtained from first-principles calculation in four ways as follows.   
The 2D DF $\epsilon(q)$ was calculated within the random-phase approximation by a first-principles supercell approach for a number of 2D crystals such as h-BN \cite{Rasmussen:2016}, phosphorene and TMDs \cite{Berkelbach:2013,Huser:2013,Rasmussen:2016}.  
In the first method, from the calculated DF the $\chi_e$ values of 2D TMDs were extracted in Ref.\cite{Berkelbach:2013} by employing the relation  $\epsilon(L)=1+4\pi \chi_e/L$  \cite{Cudazzo:2011}, where $L$ is the interlayer separation for a supercell containing two MLs of TMD, and $\epsilon(L)$ is the in-plane dielectric constant due to {\it electronic} polarization. For 2D MoS$_2$, the obtained $\chi_e$=6.6 $\AA$  \cite{Berkelbach:2013} is nearly equal to the susceptibility 6.5 $\AA$ extracted from {\it bulk} MoS$_2$ simply by taking $L=c/2$ ($c$ is the lattice constant of bulk MoS$_2$) whilst making $\epsilon(L)$ equal to the in-plane dielectric constant of bulk MoS$_2$.  When the experimental lattice constant $c$=6.66 $\AA$ \cite{Solozhenko:1995} and in-plane high-frequency dielectric constant $\epsilon=4.95$ (Ref.\cite{Geick:1966}) of bulk h-BN are put into the above relation, the susceptibility of 2D BN is estimated to be $\chi_e$=1.0 $\AA$. 
The 2D susceptibility $\chi_e$ can also be obtained by employing $\epsilon(q)=1+2\pi\chi_eq$ \cite{Cudazzo:2011} as the slope of the $\epsilon(q)$ curve \cite{Huser:2013,Rasmussen:2016} in the small-wavevector $q$ region equals the screening length $r_{eff}=2\pi\chi_e$. From Fig.~1 of Ref.\cite{Rasmussen:2016} one finds a susceptibility $\chi_e$=5.9 $\AA$ for 2D MoS$_2$ and $\chi_e$=0.8 $\AA$ for 2D hBN. 
The 2D susceptibility can also be calculated when knowing 
the screening length $r_{eff}$. For 2D hBN the calculated screening length is 7.64 $\AA$ \cite{Sohier:2017},  thus corresponding to a susceptibility $\chi_e$ of 1.216 $\AA$. 
In the fourth method, the (clamped ion) electronic energy in the unit cell is calculated for crystals in an external electric field \cite{Ferrero:2008,Ferrabone:2011}, and then the first derivative of the electronic energy with respect to the field yields the induced dipole moment per cell, namely, the induced macroscopic polarization corresponding,  for 2D hBN, to the last term in Eqs.~(\ref{bigP1}) and (\ref{bigP1z}), and the second derivative of the energy with respect to the field yields the  electronic polarizability per cell. For ML hBN an electronic polarizability 4.591 $\AA^3$ per cell (Table IV of Ref.\cite{Ferrabone:2011}) was obtained from a coupled perturbed Kohn-Sham (CPKS) calculation with the {\it ab initio} CRYSTAL09 code. Dividing this value by the area of the unit cell $s$ we find a 2D susceptibility $\chi_e$=0.848 $\AA$ which is very close to the value 0.8 $\AA$ of the second method. Similarly, using the static polarizability 7.111  $\AA^3$ per cell given in the same reference \cite{Ferrabone:2011} we find a static dielectric susceptibility (i.e., including the  lattice contribution) $\chi_0$=1.314 $\AA$. Unless otherwise stated, these 2D susceptibilities $\chi_e$ and $\chi_0$ from Ref.\cite{Ferrabone:2011} are used in this study.

Through the three expressions (\ref{a22eB}), (\ref{wto2a}) and (\ref{chiwps0}) we obtained, the three quantities of the 2D crystal, namely, the two {\it macroscopic} susceptibilities $\chi_e$, $\chi_0$ and the collective vibration frequency $\omega_0$ are related to the three {\it microscopic} quantities, $e_a$, $e_B$, $K$. Therefore, the values of the three microscopic quantities can be determined by using a set of three values $\chi_e$, $\chi_0$, $\omega_0$ calculated from first principles. Further, all the three mutually independent $a$-coefficients $a_{11}$, $a_{21}$ (or $a_{12}$), $a_{22}$ in the equation of motion (\ref{eomw1}) and polarization equation (\ref{bigP1}) can also be determined through Eqs.~(\ref{a22alf}), (\ref{a11}) and (\ref{a12chi09}). Of the four microscopic quantities $\alpha_1$, $\alpha_2$, $e_a$, $K$ on which the $a$-coefficients originally depend [refer to Eqs.~(\ref{a21}), (\ref{a22}),    (\ref{a11}), (\ref{Ke}) and (\ref{eB})],  the polarizabilities of the constituent atoms of ML hBN $\alpha_1$, $\alpha_2$ are unknown. Therefore, the adoption of the three known quantities $\chi_e$, $\chi_0$, $\omega_0$ facilitates the use of the deduced equations for in-plane motion by circumventing the two unknowns $\alpha_1$ and $\alpha_2$. We note that, although the values of $\alpha_1$, $\alpha_2$ are unknown, their sum $\alpha_1+\alpha_2$, the atomic poarizability of the unit cell, is found to fall in an interval, expressed in terms of $\chi_e$ in inequality (\ref{alfbu13}), which is obtained with a generalized  Clausius-Mossotti relation connecting the {\it microscopic} polarizabilities $\alpha_1$, $\alpha_2$ and the {\it macroscopic} dielectric susceptibility $\chi_e$ (refer to Appendix A). Taking $\chi_e=0.85 ~\AA$ \cite{Ferrabone:2011} yields $1.3084~\AA^3 \leq\alpha_1+\alpha_2\leq 1.7530~\AA^3$,  which are smaller than the total free-atom polarizability 4.0 $\AA^3$ (Appendix A).  
This interval will be used below to evaluate the LFEs on the LO phonon frequency and 2D dielectric susceptibilities   (Table~\ref{table:2} below).

Using the susceptibilities $\chi_e=0.848~\AA$ and $\chi_0=1.314~\AA$ from CPKS calculation  (Ref.\cite{Ferrabone:2011}) and the phonon frequency $\omega_0$=1371 cm$^{-1}$ calculated from first principles by the same group (Ref.\cite{Erba:2013}), we calculated the three microscopic quantities $e_a$, $e_B$, $K$ from Eqs.~(\ref{a22eB}), (\ref{wto2a}) and (\ref{chiwps0}). 
We also calculated the force constant due to LFEs $K_e$ [Eq.~(\ref{Ke})] and the LO phonon group velocity $c_l$ [Eq.~(\ref{cl})]  and present the result in Table~\ref{table:1} (upper row).  In the theoretical study by Sohier {et al}. \cite{Sohier:2017} the $\mathcal{S}$ parameter, related to $e_B$ by $\mathcal{S}=2\pi e_B^2/(\bar{m}s)$, is calculated by density-functional perturbation theory (DFPT). For 2D hBN the value $\mathcal{S}=8.4\times 10^{-2}$ eV$^2\cdot$$\AA$ \cite{Sohier:2017} so we find $e_B=2.71e$. In that study, the calculated frequency $\omega_0$ is 1387.2 cm$^{-1}$ (Table 1 of Ref.\cite{Sohier:2017}), and recall that a susceptibility $\chi_e$=1.216 $\AA$ has been already calculated above through $r_{eff}$=7.64 $\AA$ \cite{Sohier:2017}.  Now having these three values of $\chi_e$, $e_B$, $\omega_0$, again by using Eqs.~(\ref{a22eB}), (\ref{wto2a}) and (\ref{chiwps0}) we calculated the quantities $K_e$, $c_l$ as well as $\chi_0$, $e_a$, $K$, which are also listed in Table~\ref{table:1} (lower row) for comparison. First, 
we see two nearly equal values of Born charge, 2.70$e$ and 2.71$e$, the latter being also equal to another recent DFPT calculation \cite{Michel:2017} as well as the $e_B$  value  of the bulk hBN \cite{Ohba:2001}. 
According to Eq.~(\ref{a22eB}),  $e_B$ is greater than $e_a$ as $\chi_e >0$ due to EP  of ions, and the numerical result  shows that $e_B$ is four times larger than $e_a$. Also the two effective charge 
$e_a$ values are close to previous first-principles results 0.56$e$ \cite{Grad:2003} and 0.43$e$ \cite{Topsakal:2009} obtained respectively with Bader's method and L\"{o}wdin's analysis on charge transfer. 
Second, both susceptibilities $\chi_e$ and $\chi_0$ are larger in the lower row but the vibrational contributions to the static susceptibility namely the $\chi_0-\chi_e$ values are nearly equal in the two cases, which are 0.46 and 0.45 $\AA$ for the upper and lower rows, respectively. 
Third, the group velocity $c_l$ corresponds to  the slope of the LO phonon dispersion curve at $\Gamma$;  
the two values of $c_l$ are very close with a difference of only 0.38\% (Table~\ref{table:1}) but they are one order of magnitude larger than the value 1.94 km/s calculated without taking into account EP of ions \cite{Michel:2009}.  Fourth, 
 in Ref.\cite{Michel:2009} the force constant parameters are generated by making a percentage reduction of those of graphene, and from their table I we work out a spring  force constant $K$=41.86 eV/$\AA^2$ ($K'$=14.48 eV/$\AA^2$ for out-of-plane motion), which is 26\% and 30\% smaller than the force constant $K$ values 56.5 and 59.8 eV/$\AA^2$ of the present Table~\ref{table:1}. Fifth, the intrinsic oscillator frequency $\omega_0$ is determined by $K-K_e$ [Eq.~(\ref{a11})] rather than by $K$ alone, i.e., $\omega_0^2=(K-K_e)/\bar{m}$; then the ratio $K_e/K$ expresses the percentage reduction due to the LFEs,  which from Table~\ref{table:1} is  $K_e/K$=24\% and 29\%. Sixth, the two quantities $c_l/\omega_0$ and $4\pi \chi_e$ form a factor of the $k^2$ term in the expansion of $\omega_l(k)$  [Eq.~(\ref{wlok0})], and from Table~\ref{table:1} the ration is $\frac{c_l}{\omega_0}/4\pi \chi_e$=13.5\% and 9.3\%, indicating the EP of ions makes the dominant contribution. Further, this result,   together with the third point above, shows that neglecting ionic polarization such as RIM \cite{Michel:2009} cannot properly describe the LO phonon dispersion. 
Below we shall use the first row of parameters to evaluate the influence of EP and LFEs on the lattice vibrations (as is presented in Table~\ref{table:2}).

In Sec. II the electronic polarization of all the ions is included through $\alpha_{\kappa}$, and the model is referred to as PIM hereafter, in particular when comparing to the results calculated with the RIM ($\alpha_{\kappa}=0$). 
In the above calculations for Table~\ref{table:1}, therefore both the EP and LFEs have been taken into account. We now look at what happens when EP or LFEs are not included. To do this, 
we compare various lattice-dynamical quantities such as $\chi_e$, $\chi_0$, $e_B$, $\omega_t$, $\omega_l(k)$, $c_l$ obtained with the RIM and PIM when the LFEs are neglected or taken into account,  the expressions of which are listed in Table~\ref{table:2}, for the four approaches in total.  Given $\chi_e$, $\chi_0$, $\omega_0$, their expressions (\ref{a22eB}), (\ref{wto2a}) and (\ref{chiwps0}), as noted in Table~\ref{table:2},  are used to calculate the microscopic quantities such as  $e_a$, $e_B$, $K$, $\alpha_1+\alpha_2$  (see Appendix A) needed in these approaches.  
Just below the expressions in the table the specific value-substituted LO phonon dispersion $\omega_l(k)$ and the values of the other five quantities are also listed for comparison: for the PIM including the LFEs (last row), the susceptibilities $\chi_e$, $\chi_0$ and  frequency $\omega_t$ (equal to $\omega_0$) are input values taken from first-principles calculations \cite{Ferrabone:2011,Erba:2013}, while the others were calculated using this set of $\chi_e$, $\chi_0$, $\omega_0$ values, as we did in the above calculations for Table~\ref{table:1}; for the other three approaches, the quantities are transformed to depend on $e_a$ and/or $K$ as their expressions show, and are therefore calculated using the $e_a$ and $K$ parameters that are obtained with the same set of $\chi_e$, $\chi_0$, $\omega_0$ values (as was given in the first row of Table~\ref{table:1}). 
Several points can be made by comparing the results obtained with the four approaches. First, there is no electronic polarization in the RIM and therefore $\chi_e=0$. 
In the PIM with no LFEs the electronic susceptibility of ML hBN is simply $\chi_e=(\alpha_1+\alpha_2)/s$, and taking the interval we obtained above for $\alpha_1+\alpha_2$ gives $0.2417~\AA \leq\chi_e\leq 0.3239~\AA$, values that are even smaller than half the 0.85~$\AA$ obtained when the LFEs are taken into account. Second, 
in the RIM---for both cases of excluding and including the LFEs---and also in the PIM when neglecting LFEs, the Born charge reduces simply to the static charge $e_a=0.61e$, a value only 23\% of the $e_B$ that we calculated with the PIM including the LFEs. Third, the ions in motion contribute $e_B^2/[s(K-e_ae_BQ_1)]$ to the static susceptibility $\chi_0$, and with the RIM a smaller Born charge of 0.61$e$ causes this contribution to be only $\sim$0.018~$\AA$, thus underestimating the vibrational contribution by 
96\% compared to the 0.46 $\AA$ which was calculated with the PIM including the LFEs. Fourth, 
the LO phonon group velocity $c_l$ evaluated with the three simpler approaches, 1.61-1.67 km/s, is similar to the 1.94 km/s of a RIM result \cite{Michel:2009} but is one order of magnitude smaller than the 37.24 km/s given by the PIM including the LFEs.

Furthermore, the LO phonon dispersion relations given by the four quantitative expressions of Table~\ref{table:2} are compared in Fig.~\ref{fig2}, where wavevector $k$ is given in units of $\lvert \Gamma-K\rvert$ \cite{Sohier:2017}, the distance between the $\Gamma$ and $K$ points in the Brillouin zone. For the PIM without LFEs, the parameter $\gamma$ is simply taken to be the lower bound 3.8172 as the numerical calculation indicates that the use of the upper bound 5.1144, for instance, reduces the LO phonon frequencies by less than 0.08\% (not shown), a change indiscernible to the dispersion curve.  When LFEs are neglected,  the two models PIM and RIM yield very close phonon frequencies that fall in a very narrow range from 1631 to 1653 cm$^{-1}$ (upper two lines). The two curves touch each other at the $\Gamma$ point with the same tangent and slope $c_l$=1.61 km/s; the curve obtained by the RIM displays a  linear dispersion relation through the long-wavelength region $k \leq 0.15\lvert \Gamma-K\rvert$, while the PIM curve becomes flatter at larger wavevectors in the same region of $k$. 
When the LFEs are taken into account, the phonon frequencies are reduced significantly. 
For the dispersion calculated with the RIM (middle line), the dependence of frequency on wavevector remains linear, with a small slope similar to the case with no LFEs.  For the dispersion curve obtained with the PIM (solid line), in contrast, 
 owing to the EP of ions,  
a steep slope appears on the small wavevector side, corresponding to an substantially increased phonon group velocity 37.24 km/s, and as the wavevector increases the  increase of the phonon frequency becomes slower (i.e. with a smaller $c_l$), deviating significantly from its linear dependence near the $\Gamma$ point. The LO phonon dispersion of Ref.\cite{Sohier:2017}, $\omega_l=[\omega_0^2+\mathcal{S}k/(1+r_{eff}k)]^{1/2}$, 
with first-principles calculated values  $\omega_0$=1387.2 cm$^{-1}$, $\mathcal{S}=8.4\times 10^{-2}$ eV$^2\cdot$$\AA$, and $r_{eff}$=7.64 $\AA$, is also shown (dotted line) and is very close to the dispersion curve obtained with the our PIM. 
According to Eq.~(\ref{wlo2a2}), the upper limit of $\omega_l$ at a sufficiently large $k$ is $\omega_M$=1701.6 cm$^{-1}$;  
at wavevector $k=0.15\lvert \Gamma-K\rvert$, the LO phonon frequency  $\omega_l$ is 1569 cm$^{-1}$, which is quite close to its limiting value (only 7.8\% smaller). For wavevectors $k>0.02\lvert \Gamma-K\rvert$, as $\omega$ increases, the decrease of the group velocity with wavevector $k$ causes a rapid increase to the LO phonon DOS, as is clearly seen from  Fig.~\ref{fig3} [also refer to the DOS expression (\ref{doslo1})].  
Overbending is a prominent feature of the numerically calculated LO phonon dispersion of ML hBN \cite{Miyamoto:1995,Sanchez:2002,Wirtz:2003,Serrano:2007,Michel:2009,Topsakal:2009}. For instance, a maximum LO phonon frequency of 1533 cm$^{-1}$ (Ref.\cite{Wirtz:2003})  or 1570 cm$^{-1}$ (Ref.\cite{Topsakal:2009}) appears at a wavevector $\sim$0.30$\lvert \Gamma-K\rvert$.    
Evidently 
a positive slope $c_l$ is essential for the overbending of the dispersion curve $\omega_l(k)$. 
Therefore our result above indicates that due to the EP of ions, the increase of group velocity $c_l$ has enhanced the overbending substantially.

\subsection{Out-of-plane optical modes}

Values of phonon frequency $\omega_z$ \cite{Erba:2013,Topsakal:2009,Miyamoto:1995,Wirtz:2003}, 2D electronic susceptibility $\chi_e'$ and static susceptibility $\chi_0'$ \cite{Ferrabone:2011} associated with out-of-plane motion have been calculated from first principles.  Having these then the intrinsic oscillator frequency $\omega_0'$ is known through the relation (\ref{wz02chz0}). Given three quantities $\omega_0'$, $\chi_e'$, $\chi_0'$,  the three microscopic quantities namely the static effective charge $e_a'$, the Born charge $e_B'$ and the effective force constant $K'$ follow from the three equations (\ref{c22z}), (\ref{c11z}) and (\ref{chiwzs0}).  Further,  all the three mutually independent $c$-coefficients $c_{11}$, $c_{12}$ (or $c_{21}$), $c_{22}$ in the lattice equations (\ref{eomw1z}) and  (\ref{bigP1z}) are also determined by Eqs.~(\ref{c22z}), (\ref{c11z}) and (\ref{c12chi09}). 

From Ref.\cite{Ferrabone:2011} the 2D electronic susceptibility is given by $\chi_e'=0.815\AA^3/s=0.151 \AA$. Putting this into Eq.~(\ref{c22z}) gives a negative ratio $e_B^\prime/e_a^\prime=-0.21$; that is, the positive B ions carry a negative Born charge while the negative N ions move with a positive Born charge. As a result, 
the polarization induced by ionic vibrations $e_B^\prime(\mathbf{u}_1-\mathbf{u}_2)/s$ is antiparallel (parallel) to the displacements of the B (N) ions, and meanwhile the electric force exerted on the B (N) ions is antiparallel (parallel) to the macroscopic field. However the negative $e_B'$ value has no influence on the out-of-plane optical (ZO) phonon frequency as $e_B'$ enters the frequency expression via $e_B'^2$ [Eq.~(\ref{wzo2b3})].

Using the first-principles calculated values $\chi_e'=0.151~\AA$ and $\chi_0'=0.164~\AA$ (Ref.\cite{Ferrabone:2011}), we calculated $\omega_0'$, $e_a'$, $e_B'$, $K'$, $K_e'$ for two values of frequency $\omega_z$ taken from previous first-principles calculations, 836 cm$^{-1}$ (first-principles perturbation result of Ref.\cite{Erba:2013}, by the same group of Ref.\cite{Ferrabone:2011}, and also the direct method result of Ref.\cite{Miyamoto:1995}), 800 cm$^{-1}$ (DFPT value of Refs.\cite{Wirtz:2003,Topsakal:2009}), and for  
one experimental $\omega_z$ of 734 cm$^{-1}$ (Ref.\cite{Rokuta:1997}) and also for a low frequency $\omega_z$=405 cm$^{-1}$ (see Table~\ref{table:3}).  
A larger static effective charge $e_a'$, greater than  1.0$e$, occurs for out-of-plane motion than in-plane motion (compare with in-plane effective charge $e_a$ values given in Table~\ref{table:1}). Charge transfer is quite complicated in semiconductors where covalent bonds and ionic bonds coexist \cite{Phillips:1973}. For 2D hBN there is a space allowing a more electron charge distribution outside the layer plane, probably causing a larger out-of-plane charge transfer from B to N. In fact, the calculation shows that  
a smaller $e_a'$ equal to the in-plane effective charge of $0.61e$ corresponds to a very low phonon frequency 405 cm$^{-1}$ (last row of   Table~\ref{table:3}) that is unacceptably lower than the first-principles and experimental values. 
We see a negative Born charge $e_B'$ (on B ions) one order of magnitude smaller than the in-plane Born charge $e_B$,  
while the force constant due to LFEs $K_e'$ is one-third the effective spring force constant $K'$, similar to the in-plane motion case.

To examine the EP of ions and LFEs on the out-of-plane motion,   
we look at the lattice-dynamical properties $\chi_e'$, $\chi_0'$, $e_B'$, $\omega_0'$, $\omega_z$ obtained with the RIM and PIM each including or excluding the LFEs, whose expressions are listed in Table~\ref{table:4}  and also whose values obtained with a set of three first-principles values, $\chi_e'=0.151$ $\AA$, $\chi_0'=0.164$ $\AA$, $\omega_z=836$ cm$^{-1}$ (Refs.\cite{Ferrabone:2011,Erba:2013}), are given just below the corresponding expressions (the microscopic quantities $e_a'$, $e_B'$, $K'$ used in the models were calculated using the same method as above for Table~\ref{table:3}). For the adopted $\chi_e'$ value 0.151 $\AA$ the atomic polarizability of the unit cell  is found to fall in the interval, $1.9329~\AA^3 \leq\alpha_1^\prime+\alpha_2^\prime\leq 2.5792~\AA^3$ (Appendix A). In the PIM without LFEs then the dielectric susceptibility of ML hBN is  given by $\chi_e'=(\alpha_1'+\alpha_2')/s$, thus yielding $0.3571~\AA \leq\chi_e'\leq 0.4765~\AA$. 
When the LFEs are accounted for,  the dielectric susceptibility $\chi_e'$ is reduced significantly (by over 55\%)---different from the in-plane motion case where the dielectric susceptibility is increased---and the Born charge on the B ions becomes negative with a magnitude that is only 21\% of their static effective charge.  In fact,  there is a very small vibrational contribution (only 0.013 $\AA$) to the static susceptibility $\chi_0'$ due to this small value of Born charge; in contrast, when  LFEs are neglected, a large Born charge equal to the static effective charge leads to an overestimation of the vibrational contribution, which rises to a one order of magnitude higher value (0.196 $\AA$).  The  out-of-plane phonon frequency  $\omega_z$ is 836 cm$^{-1}$ with the PIM including the LFEs, which is overestimated by over 39\% when LFEs are not included.  
In the RIM  there is no EP, i.e., $\alpha_1'=\alpha_2'=0$,  which  causes $\chi_e^\prime$ to vanish and consequently an extremely large out-of-plane phonon frequency,  $\omega_z \rightarrow \infty$. In consequence the RIM fails to describe the out-of-plane vibrations.

\section{Summary and Conclusions}

We have studied long wavelength optical vibrations in ML hBN using two pairs of  equations [Eqs.~(\ref{eomw1co}), (\ref{bigP1co}), (\ref{eomw1zco}) and (\ref{bigP1zco})]
to describe the in-plane and out-of-plane lattice vibrations, respectively. 
These lattice equations, which have similar forms to Huang's equations for bulk crystals, are deduced from a microscopic dipole lattice model taking into account the LFEs and EP self-consistently. The 2D Lorentz relation connecting the macroscopic and local fields, and the use of the areal polarization, a macroscopic quantity to describe the 2D dielectric polarization, are fundamental to deducing the lattice equations from the atomic theory. From the lattice equations the expressions for the areal energy density are obtained for the in-plane and out-of-plane lattice vibrations, respectively. The three mutually independent $a$- or $c$-coefficients of the equations are expressed in terms of a set of three quantities such as the 2D electronic and static susceptibilities and the intrinsic oscillator frequency, calculated from first principles, thus making the lattice equations very useful for calculating the lattice dynamical properties.

The lattice equations are solved simultaneously with the equation of electrostatics to deduce the optical modes of the TO, LO and ZO vibrations.  
The explicit expressions have been obtained for phonon frequency, macroscopic field,  and also phonon group velocity and density of states.  The frequency expressions are found to describe the dispersion relations of previous first-principles calculations very well:  
while the TO and ZO modes are dispersionless, the LO modes have dispersion with the  frequency increasing with the wavevector.  In particular, our LO phonon dispersion relation is identical to the analytical expression of Sohier {\it et al}., and it  evidently shows that the LO and TO modes are degenerate at $\Gamma$ and split up at a finite wavevector due to the long-range macroscopic field. 
It is also shown that the finite ZO phonon frequency exists owing to the electronic polarization of ions, without which the frequency becomes infinitely large in the rigid ion model.

From the lattice equations the frequency-dependent dielectric susceptibilities are deduced for in-plane and out-of-plane lattice motion. By considering the response of the lattice to a test charge with a charge distribution asymmetric or symmetric with respect to the ML, a 2D lattice dielectric function $\epsilon(k,\omega)$ is derived [expression~(\ref{df2D})],  which is shown to be due solely to the LO vibrations, and the driven LO and ZO vibrations are further discussed. 
It is also shown that such a 2D longitudinal DF allows the LO phonon dispersion [expression~(\ref{wlo2a2})] to be rederived from $\epsilon(k,\omega)=0$,  similar to the case of bulk crystals.  
The 2D LST relation (\ref{LST2}) and frequency--susceptibility relation (\ref{wz02chz0}) are obtained for in-plane and out-of-plane motion, respectively, connecting the phonon frequencies to the 2D dielectric functions or susceptibilities, which are very useful for quantifying the effects of the macroscopic field on the phonon frequencies.  
 
The analytical expressions have been applied to study the lattice dynamical properties of ML hBN. 
For the in-plane motion, two sets of three quantities from two independent first-principles calculations, one set of $\chi_e$, $\chi_0$, $\omega_0$ from Ref.\cite{Ferrabone:2011} and the other set of $\chi_e$, $e_B$, $\omega_0$  from Ref.\cite{Sohier:2017}, are used as parameters and very close results are obtained for the calculated properties such as the LO phonon dispersion relation, the effective spring force constant  and the  vibrational contribution to the static susceptibility. 
To evaluate the LFEs the unit-cell atomic polarizability is used, which, given a dielectric susceptibility, is found to be limited in an interval using the 2D Clausius-Mossotti relation.    
The LFEs and electronic polarization of ions should be included simultaneously, but otherwise neglecting either or both causes large discrepancies to the calculated properties: the phonon frequency at $\Gamma$ is overestimated by 15\%-19\%, whereas the Born charge, the LO phonon group velocity and the vibrational contribution to the static susceptibility are underestimated by 77\%, 96\%, 96\%, respectively. With no ionic EP or LFEs, the LO modes display very small linear dispersion, almost flat in the long wavelength region, which is distinct from the LO phonon dispersion calculated after accounting for both EP and LFEs. 
For out-of-plane motion, using $\chi_e'$, $\chi_0'$, $\omega_z$ calculated mainly from Refs.\cite{Ferrabone:2011,Erba:2013} 
as parameters, 
 the static effective charge is found to be different from that for in-plane motion reflecting anisotropy of the 3D charge density distribution. 
Further, the positive B ions carry a negative Born charge while the negative N ions move with a positive Born charge. Similar to the in-plane motion case, the RIM can not  properly describe the out-of-plane vibrations, which, for instance, yields a four times larger Born charge and infinitely large phonon frequency. 
When the EP of ions is included, the LFEs have significantly reduced the 2D dielectric susceptibility and the Born charge (by 60\% and 80\%, respectively), different from the in-plane motion case where the dielectric susceptibility and Born charge are both increased substantially (by two and three times, respectively).


\begin{acknowledgments}
We acknowledge support from the Natural Science Research Funds (Nos. 419080500175 \& 419080500260) of Jilin University. 
\end{acknowledgments}


\appendix 

\section{The generalized Clausius-Mossotti relations and intervals of the total atomic polarizability in ML hBN}   

In bulk diatomic crystals the electronic susceptibility $\chi$ relates to the atomic polarizabilities of constituent atoms $\alpha_1$ and $\alpha_2$ by the Clausius-Mossotti relation \cite{Clausius:1879,Mossotti:1850}, 
\begin{equation}
\chi=\frac{(\alpha_1+\alpha_2)/v_a}{1-4\pi(\alpha_1+\alpha_2)/(3v_a)}.   
\label{clausius3d}
\end{equation}

There is no such a simple relation for 2D hBN. Substituting Eqs.~(\ref{bigD}) and (\ref{a22alf}) into Eq.~(\ref{a22}), $\chi_e$, the in-plane electronic susceptibility of ML hBN, is given by  \cite{Mikhailov:2013,Della:2016}
\begin{equation}
\chi_e=\frac{(\alpha_1+\alpha_2)+2\alpha_1\alpha_2(Q_1-Q_0)}{s[1-(\alpha_1+\alpha_2)Q_0-\alpha_1\alpha_2(Q_1^2-Q_0^2)]}.   
\label{clausius1}
\end{equation}

If $Q_0=Q_1$, then the 2D susceptibility of the dielectric $\chi_e$ relates to the total atomic polarizability of the unit cell $\alpha_1+\alpha_2$ by a simple relation, 
\begin{equation}
\chi_e=\frac{(\alpha_1+\alpha_2)/s}{1-(\alpha_1+\alpha_2)Q_{0,1}},    
\label{clausius2}
\end{equation}
which is similar to the Clausius-Mossotti relation (\ref{clausius3d}) for bulk materials.  In fact in ML hBN $Q_0 \neq Q_1$ (Sec. II) so the sum $\alpha_1+\alpha_2$ can not be determined from the generalized Clausius-Mossotti relation (\ref{clausius1}); adding the Born charge expression (\ref{eB}) does not help because $e_B$ and $\chi_e$ are related through Eq.~(\ref{a22eB}) and not independent of each other. 

Interestingly, when knowing $\chi_e$ the upper and lower bounds of the unit-cell atomic polarizability $\alpha_1+\alpha_2$ can be determined from expression (\ref{clausius1}). 
Let $\alpha=\alpha_1+\alpha_2$. Then Eq.~(\ref{clausius1}) can be transformed into a quadratic equation in variable $\alpha_1$, 
\begin{equation}
(Q_1-Q_0)\big(Q_1+Q_0+\frac{2}{\chi_es}\big)(\alpha_1^2-\alpha\alpha_1)+1-\alpha\big(Q_0+\frac{1}{\chi_es}\big)=0,     
\label{alfbo1}
\end{equation}
with the discriminant $\Delta_{\alpha}$, 
\begin{align}
\Delta_{\alpha}&=(Q_1-Q_0)\big(Q_1+Q_0+\frac{2}{\chi_es}\big)\Big\{(Q_1-Q_0)\big(Q_1+Q_0 \big.\Big.
\nonumber \\ &\qquad {}
\Big.\big.+\frac{2}{\chi_es}\big)\alpha^2-4\big[1-\alpha(Q_0+\frac{1}{\chi_es})\big]\Big\}   \nonumber \\  
& =(Q_1-Q_0)^2\big(Q_1+Q_0+\frac{2}{\chi_es}\big)^2 \Big(\alpha+\frac{2}{Q_1-Q_0}\Big) 
\nonumber \\ &\qquad {}
\Big(\alpha-\frac{2}{Q_1+Q_0+\frac{2}{\chi_es}}\Big).   
\label{alfdel1}
\end{align}

Eq.~(\ref{alfbo1}) has two real roots only when $\Delta_{\alpha}\geq 0$. Therefore it follows from the $\Delta_{\alpha}$ expression (\ref{alfdel1}) that either 
\begin{equation}
\alpha\geq\frac{2}{Q_1+Q_0+\frac{2}{\chi_es}},  
\label{alfbu1}
\end{equation}
or 
\begin{equation}
\alpha\leq-\frac{2}{Q_1-Q_0}.  
\label{alfbu2}
\end{equation}
For ML hBN ($Q_1>Q_0$) the latter inequality leads to only negative $\alpha$ and is dropped. 
Now one finds that $\alpha_1$ and $\alpha_2$ are given by 
\begin{equation}
\alpha_{1,2}=\frac{1}{2}\Big[\alpha\pm\sqrt{\Big(\alpha+\frac{2}{Q_1-Q_0}\Big) \Big(\alpha-\frac{2}{Q_1+Q_0+\frac{2}{\chi_es}}\Big)}~\Big]; 
\label{alf1root}
\end{equation}
if $\alpha_1$ takes the plus square root then $\alpha_2$ takes the minus square root, and vice versa. 
For $\alpha_1$ and $\alpha_2$ to be positive, one further requires that the above square root is not greater than $\alpha$ and obtains 
\begin{equation}
\alpha\leq\frac{1}{Q_0+\frac{1}{\chi_es}}. 
\label{alfbu3}
\end{equation}

Combining the inequalities (\ref{alfbu1}) and (\ref{alfbu3}) then restricts the atomic polarizability in the unit cell for in-plane polarization to the following interval, 
\begin{equation}
\frac{2}{Q_1+Q_0+\frac{2}{\chi_es}}\leq\alpha_1+\alpha_2\leq\frac{1}{Q_0+\frac{1}{\chi_es}}. 
\label{alfbu13}
\end{equation}
When the 2D susceptibility $\chi_e=0.85 ~\AA$ \cite{Ferrabone:2011} is 
used, for instance, one finds $1.3084~\AA^3 \leq\alpha_1+\alpha_2\leq 1.7530~\AA^3$,  
which is only a narrow range of 0.44~$\AA^3$. 

For the vertical polarization (parallel to $\mathbf{e}_z$),  the electronic susceptibility  $\chi_e^\prime$ can be expressed in terms of the atomic polarizabilities $\alpha_1^\prime$, $\alpha_2^\prime$ as (refer to Sec. II above and also Ref.\cite{Mikhailov:2013})
\begin{equation}
\chi_e^\prime=\frac{(\alpha_1^\prime+\alpha_2^\prime)-4\alpha_1^\prime\alpha_2^\prime(Q_1-Q_0)}{s[1+2(\alpha_1^\prime+\alpha_2^\prime)Q_0-4\alpha_1^\prime\alpha_2^\prime(Q_1^2-Q_0^2)]}.    
\label{clausiusz1}
\end{equation}

Following the preceding derivation we find that the atomic polarizability of the unit  cell falls in the interval, 
\begin{equation}
\frac{1}{\frac{1}{\chi_e^\prime s}-2Q_0}\leq\alpha_1^\prime+\alpha_2^\prime\leq\frac{1}{Q_1-Q_0}. 
\label{alfbuz13}
\end{equation}

Taking the 2D susceptibility $\chi_e^\prime=0.151 ~\AA$ \cite{Ferrabone:2011}, one finds $1.9329~\AA^3 \leq\alpha_1^\prime+\alpha_2^\prime\leq 2.5792~\AA^3$. 
Mathematically, the solution $\alpha_1^\prime+\alpha_2^\prime \geq 1/(\frac{1}{\chi_e^\prime s}-Q_1-Q_0)$ is also permitted, but yielding $\alpha_1^\prime+\alpha_2^\prime \geq 7.3 ~\AA^3$ for ML hBN, which are much larger than the total polarizability 4.0 $\AA^3$ of the free atoms B and N (the polarizabilities of free atoms B and N are 3.038 and 1.097 $\AA^3$, respectively \cite{PScollect:2006}). Using the linear volume-polarizability relationship \cite{Brinck:1993}, the polarizability of a constituent atom $\alpha_i$ in a crystal can be estimated by $\alpha_i=(v_i/v_i^0)\alpha_i^0$ \cite{Tkatchenko:2009,Hod:2012}, where $v_i$ is the effective volume of the atom in the crystal, and $v_i^0$ and $\alpha_i^0$ are the free-atom volume and polarizability, respectively. Numerical calculations have shown that the polarizabilities of the B and N atoms in bulk h-BN are reduced compared to their free-atom polarizabilities \cite{Tkatchenko:2009,Hod:2012}. Anyhow the values $\alpha_1^\prime+\alpha_2^\prime \geq 7.3 ~\AA^3$  are unacceptably large to be considered. 


Compared to the their free-atom polarizability 4.0 $\AA^3$, the atoms B and N have a significantly reduced total polarizability  after forming the 2D crystal hBN. There is a slightly larger atomic polarizability in the vertical direction than in the layer plane, i.e., $\alpha_1^\prime+\alpha_2^\prime >\alpha_1+\alpha_2$ (with a 0.6 $\AA^3$ difference in upper bound), as the constituent B and N atoms are close packed on the plane causing the electron cloud to be more easily deformed and polarized along the vertical direction. 
If LFEs are neglected then the dielectric susceptibilities become $\chi_e=(\alpha_1+\alpha_2)/s$, and $\chi_e^\prime=(\alpha_1^\prime+\alpha_2^\prime)/s$, leading to $\chi_e^\prime > \chi_e$.  Therefore the first-principles result $\chi_e > \chi_e^\prime$ as shown in Ref.\cite{Ferrabone:2011} is attributed to the LFEs.

We note that $\chi_e$ and $\chi_e^\prime$ are susceptibilities due to polarizable atoms or ions, and thus they also occur in non-ionic 2D crystals such as graphene. For graphene expressions (\ref{clausius1}) and (\ref{clausiusz1}) for $\chi_e$ and $\chi_e^\prime$ are still applicable but become simpler considering there are two C atoms in a unit cell, namely $\alpha_1=\alpha_2$ and $\alpha_1^\prime=\alpha_2^\prime$.

\section{Proof of $a_{12}=a_{21}$ and $c_{12}=c_{21}$ from macroscopic theory and areal energy density in 2D hBN}   

We first consider in-plane motion. Let the ions have a displacement in the layer plane, $\mathbf{w}(\boldsymbol{\rho})=\mathbf{w}_{\boldsymbol{\rho}}(\boldsymbol{\rho})$, in the presence of a point charge $e_{\nu}$ fixed at ($\boldsymbol{\rho}_0$,0), and let us find the electric field in the monolayer (in-plane component), $\mathbf{E}_{\boldsymbol{\rho}}(\boldsymbol{\rho},0)$.  The polarization $\mathbf{P}$ is  $\mathbf{P(\mathbf{r})}=[a_{21}\mathbf{w}(\boldsymbol{\rho})+a_{22}\mathbf{E}_{\boldsymbol{\rho}}(\boldsymbol{\rho},0)+c_{22}\mathbf{E}_z(\boldsymbol{\rho},0)]\delta(z)$ upon using Eqs.~(\ref{bigP1co}) and (\ref{bigP1zco}), and the Poisson equation is given by  
\begin{align}
&\nabla^2\phi(\boldsymbol{\rho},z)=4\pi\{\nabla_{\boldsymbol{\rho}}\cdot[a_{21}\mathbf{w}(\boldsymbol{\rho})+a_{22}\mathbf{E}_{\boldsymbol{\rho}}(\boldsymbol{\rho},0)]\delta(z)  
\nonumber \\  &\qquad {}
+c_{22}E_z(\boldsymbol{\rho},0)\delta'(z)-e_{\nu}\delta(\boldsymbol{\rho}-\boldsymbol{\rho}_0)\delta(z)\}. 
\label{poiapp1}
\end{align}

Expanding $\phi(\boldsymbol{\rho},z)$, $\mathbf{w}(\boldsymbol{\rho})$ and $\delta(\boldsymbol{\rho}-\boldsymbol{\rho}_0)$,
\begin{equation}
\phi(\boldsymbol{\rho},z)=\sum_{\mathbf{k}}\varphi_{\mathbf{k}}(z)e^{i\mathbf{k}\cdot\boldsymbol{\rho}}, 
\label{phixapp1}
\end{equation}
\begin{equation}
\mathbf{w}(\boldsymbol{\rho})=\sum_{\mathbf{k}}\mathbf{w}_{\mathbf{k}}e^{i\mathbf{k}\cdot\boldsymbol{\rho}}, 
\label{wxapp1}
\end{equation}
\begin{equation}
\delta(\boldsymbol{\rho}-\boldsymbol{\rho}_0)=\frac{1}{A}\sum_{\mathbf{k}}e^{i\mathbf{k}\cdot(\boldsymbol{\rho}-\boldsymbol{\rho}_0)},
\label{delroxapp1}
\end{equation}
$A$ being the sample area, Eq.~(\ref{poiapp1}) is then transformed to  
\begin{align}
&(\nabla_{\boldsymbol{\rho}}^2+\frac{\partial^2}{\partial z^2})\sum_{\mathbf{k}}\varphi_{\mathbf{k}}(z)e^{i\mathbf{k}\cdot\boldsymbol{\rho}}=4\pi\sum_{\mathbf{k}}\left [\delta(z)(a_{21}i\mathbf{k}\cdot\mathbf{w}_{\mathbf{k}}
\right. 
\nonumber \\ &\qquad {} \left. +a_{22}k^2\varphi_{\mathbf{k}}(0))  
-\delta'(z)c_{22}\varphi'_{\mathbf{k}}(0)\right ] e^{i\mathbf{k}\cdot\boldsymbol{\rho}}  
\nonumber \\ &\qquad {}  -\frac{4\pi e_{\nu}}{A}\sum_{\mathbf{k}}e^{i\mathbf{k}\cdot(\boldsymbol{\rho}-\boldsymbol{\rho}_0)}\delta(z). 
\label{poiapp2}
\end{align}

Expanding $\varphi_{\mathbf{k}}(z)$ 
\begin{equation}
\varphi_{\mathbf{k}}(z)=\int_{-\infty}^{\infty}\hat{\varphi}_{\mathbf{k}}(q)e^{iqz}dq,
\label{vphxapp1}
\end{equation}
and $\delta(z)$ as Eq.~(\ref{delx}), one finds $\hat{\varphi}_{\mathbf{k}}(q)$ from  Eq.~(\ref{poiapp2}), 
\begin{align}
\hat{\varphi}_{\mathbf{k}}(q)&=\frac{2}{k^2+q^2}\left[-(a_{21}i\mathbf{k}\cdot\mathbf{w}_{\mathbf{k}}+a_{22}k^2\varphi_{\mathbf{k}}(0))  \right. \nonumber \\  &\qquad {}
\left.+\frac{e_{\nu}}{A}e^{-i\mathbf{k}\cdot\boldsymbol{\rho}_0}+c_{22}\varphi'_{\mathbf{k}}(0)iq\right]. 
\label{vphkqapp1}
\end{align}

We note that from the above expansions one has reality conditions $\varphi_{-\mathbf{k}}(z)=\varphi^*_{\mathbf{k}}(z)$, $\mathbf{w}_{-\mathbf{k}}=\mathbf{w}^*_{\mathbf{k}}$ and $\hat{\varphi}_{-\mathbf{k}}(-q)=\hat{\varphi}^*_{\mathbf{k}}(q)$ considering that quantities such as $\phi(\boldsymbol{\rho},z)$, $\mathbf{w}(\boldsymbol{\rho})$ are real.  

Taking $z=0$ in Eq.~(\ref{vphxapp1}) and integrating $\hat{\varphi}_{\mathbf{k}}(q)$ over $q$ one finds   
\begin{equation}
\varphi_{\mathbf{k}}(0)=\frac{2\pi}{k(1+2\pi a_{22}k)}\left(-a_{21}i\mathbf{k}\cdot\mathbf{w}_{\mathbf{k}}+\frac{e_{\nu}}{A}e^{-i\mathbf{k}\cdot\boldsymbol{\rho}_0}\right), 
\label{vphz0app1}
\end{equation}
and then the in-plane component of the electric field in the monolayer,
\begin{align}
\mathbf{E}_{\boldsymbol{\rho}}(\boldsymbol{\rho},0)&=-\sum_{\mathbf{k}}\frac{2\pi\mathbf{k}}{k(1+2\pi a_{22}k)}\left(a_{21}\mathbf{k}\cdot\mathbf{w}_{\mathbf{k}} \right. \nonumber \\  &\qquad {} 
\left.+\frac{ie_{\nu}}{A}e^{-i\mathbf{k}\cdot\boldsymbol{\rho}_0}\right)e^{i\mathbf{k}\cdot\boldsymbol{\rho}}, 
\label{Eroz0app1}
\end{align}
where the latter sum represents the field due to the point charge (including dielectric effects through $a_{22}$), which is zero at the charge site. 
Eq.~(\ref{Eroz0app1}) will be used below to calculate the work required to displace the ions or the charge. 

We have obtained Eq.~(\ref{a12a21}), namely, $a_{12}=a_{21}$, from the microscopic dipole lattice model, and here we show that from the viewpoint of macroscopic theory it follows from the principle of energy conservation.  
Place the point charge $e_{\nu}$ at the origin, whilst setting the ions in the configuration $\mathbf{w}(\boldsymbol{\rho})=0$, and consider the following cycle \cite{Born:1954}: (a) keeping the charge at the origin, displace the ions horizontally into an irrotational configuration $\mathbf{w}(\boldsymbol{\rho})=\nabla_{\boldsymbol{\rho}}\psi(\boldsymbol{\rho})$, according to 
\begin{equation}
\mathbf{w}(\boldsymbol{\rho})=\xi\nabla_{\boldsymbol{\rho}}\psi(\boldsymbol{\rho}), 
\label{wroxiap}
\end{equation}
by increasing $\xi$ from 0 to 1, $\psi(\boldsymbol{\rho})$ being a nonzero arbitrary scalar field; (b) keeping the ions at $\mathbf{w}(\boldsymbol{\rho})=\nabla_{\boldsymbol{\rho}}\psi(\boldsymbol{\rho})$, displace the charge to $\Delta \boldsymbol{\rho}$; (c) fixing the charge at $\Delta \boldsymbol{\rho}$, reverse process (a), i.e., by decreasing $\xi$ from 1 to 0  according to Eq.~(\ref{wroxiap}); (d) move the charge back to the origin to complete the cycle. 

$-[a_{11}\mathbf{w}(\boldsymbol{\rho})+a_{12}\mathbf{E}_{\boldsymbol{\rho}}(\boldsymbol{\rho},0)]\cdot \Delta \mathbf{w}(\boldsymbol{\rho})$ is the work per unit area required to change $\mathbf{w}(\boldsymbol{\rho})$ to $\mathbf{w}(\boldsymbol{\rho})+\Delta \mathbf{w}(\boldsymbol{\rho})$, and total work expended on the ionic system for the configuration change is 
\begin{equation}
-\int [a_{11}\mathbf{w}(\boldsymbol{\rho})+a_{12}\mathbf{E}_{\boldsymbol{\rho}}(\boldsymbol{\rho},0)]\cdot \Delta \mathbf{w}(\boldsymbol{\rho})d\boldsymbol{\rho}. 
\label{dewpaap1}
\end{equation}
From Eq.~(\ref{wroxiap}) 
$\Delta\mathbf{w}(\boldsymbol{\rho})=\Delta\xi\nabla_{\boldsymbol{\rho}}\psi(\boldsymbol{\rho})$ follows, and then insert expansion $\psi(\boldsymbol{\rho})=\sum_{\mathbf{k}}\psi_{\mathbf{k}}e^{i\mathbf{k}\cdot\boldsymbol{\rho}}$ into 
$\mathbf{w}(\boldsymbol{\rho})$ [Eq.~(\ref{wroxiap})] and $\Delta\mathbf{w}(\boldsymbol{\rho})$. Comparing the former expansion with Eq.~(\ref{wxapp1}) one  finds $\mathbf{w}_{\mathbf{k}}$ for Eq.~(\ref{Eroz0app1}),
$\mathbf{w}_{\mathbf{k}}=\xi i\mathbf{k}\psi_{\mathbf{k}}$.
Inserting these expansions of $\mathbf{w}(\boldsymbol{\rho})$ and $\Delta\mathbf{w}(\boldsymbol{\rho})$ together with Eq.~(\ref{Eroz0app1}) into the above expression (\ref{dewpaap1}), and integrating over $\xi$ from 0 to 1, one obtains the work done during process (a), 
\begin{align}
W_a&=\sum_{\mathbf{k}}\left[\frac{1}{2}A\left(\frac{2\pi a_{12}a_{21}k}{1+2\pi a_{22}k}-a_{11}\right)\psi_{\mathbf{k}}  \right. \nonumber \\ &\qquad {} 
\left. +\frac{2\pi a_{12}e_{\nu}}{k(1+2\pi a_{22}k)}\right]k^2\psi_{-\mathbf{k}}.  
\label{workpaap1}
\end{align}
 
The field acting on charge $e_{\nu}$ during process (b) is given by setting $\boldsymbol{\rho}=0$ (as $\Delta \boldsymbol{\rho}$ is small) and $\mathbf{w}_{\mathbf{k}}=i\mathbf{k}\psi_{\mathbf{k}}$ in the first term of $\mathbf{E}_{\boldsymbol{\rho}}(\boldsymbol{\rho},0)$ [Eq.~(\ref{Eroz0app1})], and the work expended during the process is 
\begin{equation}
W_b=2\pi ie_{\nu}a_{21}\sum_{\mathbf{k}} \frac{k\psi_{\mathbf{k}}\mathbf{k}}{1+2\pi a_{22}k}\cdot \Delta \boldsymbol{\rho}.
\end{equation}
 
Process (c) is the reverse of process (a) except for the altered position of charge $e_{\nu}$. Therefore after reversing the sign of  expression (\ref{workpaap1}) and multiplying its latter summands by a factor $e^{-i\mathbf{k}\cdot\Delta \boldsymbol{\rho}}$ we find the work expended during the process,
\begin{align}
W_c&=-\sum_{\mathbf{k}}\left[\frac{1}{2}A\left(\frac{2\pi a_{12}a_{21}k}{1+2\pi a_{22}k}-a_{11}\right)\psi_{\mathbf{k}}  \right. \nonumber \\ &\qquad {} \left. 
+\frac{2\pi a_{12}e_{\nu}}{k(1+2\pi a_{22}k)}e^{-i\mathbf{k}\cdot\Delta \boldsymbol{\rho}}\right]k^2\psi_{-\mathbf{k}}.  
\label{workpcap1}
\end{align}

During process (d) $\mathbf{w}(\boldsymbol{\rho})=0$, and according to  Eq.~(\ref{Eroz0app1}) no field acts on charge $e_{\nu}$, and therefore no work is needed to restore the charge to the origin, $W_d=0$. 

As $W_a+W_b+W_c+W_d=0$, one finds
\begin{align}
&2\pi a_{12}e_{\nu}\sum_{\mathbf{k}}\frac{k\psi_{-\mathbf{k}}}{1+2\pi a_{22}k}\left(1-e^{-i\mathbf{k}\cdot\Delta \boldsymbol{\rho}}\right)
\nonumber \\ &\qquad {} 
+2\pi ia_{21}e_{\nu}\sum_{\mathbf{k}}\frac{k\psi_{\mathbf{k}}\mathbf{k}}{1+2\pi a_{22}k}\cdot \Delta \boldsymbol{\rho}=0. 
\label{worksap1}
\end{align}

Upon making $e^{-i\mathbf{k}\cdot\Delta \boldsymbol{\rho}}=1-i\mathbf{k}\cdot\Delta \boldsymbol{\rho}$, and changing the index of summation $\mathbf{k}$ to $-\mathbf{k}$ for the former summation, one immediately obtains $a_{12}=a_{21}$.

The relation $a_{12}=a_{21}$ allows us to define an energy density (energy per unit area) $u_p$ associated with the in-plane optical vibrations, as a function of $\mathbf{w}_{\boldsymbol{\rho}}(\boldsymbol{\rho})$ and $\mathbf{E}_{\boldsymbol{\rho}}(\boldsymbol{\rho},0)$,  
\begin{equation}
u_h=-\frac{1}{2}\left[a_{11}\mathbf{w}^2_{\boldsymbol{\rho}}(\boldsymbol{\rho})+2a_{12}\mathbf{w}_{\boldsymbol{\rho}}(\boldsymbol{\rho})\cdot\mathbf{E}_{\boldsymbol{\rho}}(\boldsymbol{\rho},0)+a_{22}\mathbf{E}^2_{\boldsymbol{\rho}}(\boldsymbol{\rho},0)\right].  
\label{endenuh}
\end{equation}
Inserting this $u_h$ expression into $\ddot{\mathbf{w}}_{\boldsymbol{\rho}}(\boldsymbol{\rho})=-\nabla _{\mathbf{w}}u_h$, and $\boldsymbol{\mathcal{P}}(\boldsymbol{\rho})=-\nabla _{\mathbf{E}}u_h$, where $\nabla _{\mathbf{w}}=(\frac{\partial}{\partial w_x},\frac{\partial}{\partial w_y})$, for instance, and the vector subscripts $\mathbf{w}$ and $\mathbf{E}$ of del $\nabla$ are shorthand notations for in-plane vectors $\mathbf{w}_{\boldsymbol{\rho}}(\boldsymbol{\rho})$ and $\mathbf{E}_{\boldsymbol{\rho}}(\boldsymbol{\rho},0)$, one  rederives the lattice equations (\ref{eomw1co}) and (\ref{bigP1co}). 
We note that,   
similar to the bulk polar crystal case \cite{Born:1954}, it is not {\it a priori} obvious that an energy density of this simple form should exist for ML hBN, in particular when considering $\mathbf{E}$ is the macroscopic field -- not simply an externally applied field.  

Now we consider out-of-plane motion. Let the ions have a displacement perpendicular to the layer plane, $\mathbf{w}(\boldsymbol{\rho})=\mathbf{w}_z(\boldsymbol{\rho})$, in the presence of a point charge $e_{\nu}$ placed at (0,$z_0$), and let us find the $z$-component  of the field, $\mathbf{E}_z(\boldsymbol{\rho},z)$. Now the polarization $\mathbf{P}$ is  $\mathbf{P(\mathbf{r})}=[c_{21}\mathbf{w}(\boldsymbol{\rho})+c_{22}\mathbf{E}_z(\boldsymbol{\rho},0)+a_{22}\mathbf{E}_{\boldsymbol{\rho}}(\boldsymbol{\rho},0)]\delta(z)$ upon using Eqs.~(\ref{bigP1co}) and (\ref{bigP1zco})], and the Poisson equation is given by  
\begin{align}
&\nabla^2\phi(\boldsymbol{\rho},z)=4\pi\{[c_{21}\mathbf{w}(\boldsymbol{\rho})+c_{22}\mathbf{E}_z(\boldsymbol{\rho},0)]\delta'(z)  \nonumber \\ &\qquad {}  
+ a_{22}\nabla_{\boldsymbol{\rho}}\cdot\mathbf{E}_{\boldsymbol{\rho}}(\boldsymbol{\rho},0)\delta(z)-e_{\nu}\delta(\boldsymbol{\rho})\delta(z-z_0)\}. 
\label{poiapz1}
\end{align}

Expanding $\delta(z)$ [Eq.~(\ref{delx})] and the other spatially varying quantities  exactly as in the in-plane motion case [Eqs.~(\ref{phixapp1}), (\ref{wxapp1}), (\ref{delroxapp1}) and (\ref{vphxapp1})], one finds from the Poisson equation $\hat{\varphi}_{\mathbf{k}}(q)$, 
\begin{align}
\hat{\varphi}_{\mathbf{k}}(q)&=\frac{2}{k^2+q^2}\left[-(c_{21}\mathbf{w}_{\mathbf{k}}\cdot\mathbf{e}_z-c_{22}\varphi'_{\mathbf{k}}(0))iq \right. \nonumber \\  &\qquad {} 
\left. +\frac{e_{\nu}}{A}e^{-iqz_0}-a_{22}k^2\varphi_{\mathbf{k}}(0)\right]. 
\label{vphkqapz1}
\end{align}

Taking $z=0$ in Eq.~(\ref{vphxapp1}) and integrating $\hat{\varphi}_{\mathbf{k}}(q)$ over $q$ one finds   
\begin{equation}
\varphi_{\mathbf{k}}(0)=\frac{2\pi e_{\nu}}{Ak(1+2\pi a_{22}k)}e^{-k\lvert z_0\rvert}.   
\label{vphz0apz1}
\end{equation}

Differentiating Eq.~(\ref{vphxapp1}) with respect to $z$ and substituting  Eq.~(\ref{vphkqapz1}) for $\hat{\varphi}_{\mathbf{k}}(q)$ and then integrating over $q$, one obtains  
\begin{align}
\varphi'_{\mathbf{k}}(z)&=2\pi \Big\{2(c_{21}\mathbf{w}_{\mathbf{k}}\cdot\mathbf{e}_z-c_{22}\varphi'_{\mathbf{k}}(0))\Big[\delta(z)-\frac{1}{2}ke^{-k\lvert z\rvert}\Big] \Big .
\nonumber \\ &\qquad {} -\frac{e_{\nu}}{A}\sgn(z-z_0)e^{-k\lvert z-z_0\rvert} 
\nonumber \\ &\qquad {}  \Big. +a_{22}k^2\varphi_{\mathbf{k}}(0)\sgn(z)e^{-k\lvert z\rvert}\Big\}. 
\label{vph1kzapz1}
\end{align}

To find $\varphi'_{\mathbf{k}}(0)$ we approach $\delta(z)$ with $\delta_{\varepsilon}(z)$ where $\varepsilon$ is a small thickness \cite{Michel:2009,Sohier:2017} and let $\Lambda=\delta_{\varepsilon}(0)$.  Thus from Eq.~(\ref{vph1kzapz1}) one finds 
\begin{align}
\varphi'_{\mathbf{k}}(0)&=\frac{2\pi}{1+4\pi c_{22}(\Lambda-k/2)}\left[2c_{21}(\Lambda-\frac{k}{2})\mathbf{w}_{\mathbf{k}}\cdot\mathbf{e}_z   \right. 
\nonumber \\ &\qquad {} \left. +\frac{e_{\nu}}{A}\sgn(z_0)e^{-k\lvert z_0\rvert}\right].   
\label{vph1kz0apz1}
\end{align}

Having $\varphi_{\mathbf{k}}(0)$  and $\varphi'_{\mathbf{k}}(0)$ [Eqs.~(\ref{vphz0apz1}) and (\ref{vph1kz0apz1})], now insert them  into Eq.~(\ref{vph1kzapz1}) for $\varphi'_{\mathbf{k}}(z)$, from which one obtains $\mathbf{E}_z(\boldsymbol{\rho},z)$ via $\mathbf{E}_z(\boldsymbol{\rho},z)=-\mathbf{e}_z\sum_{\mathbf{k}}\varphi'_{\mathbf{k}}(z)e^{i\mathbf{k}\cdot\boldsymbol{\rho}}$,  
\begin{align}
\mathbf{E}_z(\boldsymbol{\rho},z)&= \sum_{\mathbf{k}}\Big\{ \frac{2\pi}{1+4\pi c_{22}(\Lambda-k/2)}\Big(-2c_{21}\mathbf{w}_{\mathbf{k}} \Big. \Big. \nonumber \\  &\qquad {} 
\Big.+4\pi c_{22}\frac{e_{\nu}}{A}\sgn(z_0)e^{-k\lvert z_0\rvert}\mathbf{e}_z\Big)\big[\delta(z)-\frac{1}{2}ke^{-k\lvert z\rvert}\big]   
\nonumber \\ 
&\qquad {}  \Big. +\frac{2\pi e_{\nu}}{A}\big[-\frac{2\pi a_{22}k}{1+2\pi a_{22}k}\sgn(z)e^{-k(\lvert z\rvert +\lvert z_0\rvert)}  \big. 
\nonumber \\  &\qquad {} 
\Big.\big.+\sgn(z-z_0)e^{-k\lvert z-z_0\rvert}\big]\mathbf{e}_z \Big\}e^{i\mathbf{k}\cdot\boldsymbol{\rho}}. 
\label{vph1kzapz2}
\end{align}
 
Let $z_0>0$ in what follows. Taking $z=0$ yields the $z$-component of the field in the ML,  
\begin{align}
\mathbf{E}_z(\boldsymbol{\rho},0)&=-2\pi\sum_{\mathbf{k}}\frac{1}{1+4\pi c_{22}(\Lambda-k/2)}\left[2c_{21}(\Lambda-\frac{k}{2})\mathbf{w}_{\mathbf{k}} \right. \nonumber \\  &\qquad {}
\left. +\frac{e_{\nu}}{A}e^{-k z_0}\mathbf{e}_z\right]e^{i\mathbf{k}\cdot\boldsymbol{\rho}},     
\label{Ez0apz2}
\end{align}
while taking $z=z_0$ and $\boldsymbol{\rho}=0$ one finds the $z$-component of the field acting on the point charge $e_{\nu}$,  
\begin{align}
&\mathbf{E}_z(0,z_0)=-2\pi \sum_{\mathbf{k}}\Big\{ \frac{k}{1+4\pi c_{22}(\Lambda-k/2)}\Big(-c_{21}\mathbf{w}_{\mathbf{k}}  \Big. \Big . 
\nonumber \\ &\qquad {}  \Big. \Big. +2\pi c_{22}\frac{e_{\nu}}{A}e^{-k z_0}\mathbf{e}_z\Big)
+\frac{e_{\nu}}{A}\frac{2\pi a_{22}k}{1+2\pi a_{22}k}e^{-k z_0}\mathbf{e}_z \Big\}e^{-k z_0}, 
\label{Ezz0apz3}
\end{align}
where the sum of the terms containing $e_{\nu}$ ($\mathbf{w}_{\mathbf{k}}$) represents the field due to the point charge (ionic displacements), denoted by $\mathbf{E}_{z,e_{\nu}}(0,z_0)$ ($\mathbf{E}_{z,w}(0,z_0)$) for simplicity.  

Now we show that  $c_{12}=c_{21}$ [Eq.~(\ref{c12z})]  follows from the principle of energy conservation. Place the point charge $e_{\nu}$ above the ML at a point $P$ ($\boldsymbol{\rho}_0=0$,$z_0$), while keeping the ions in the configuration $\mathbf{w}(\boldsymbol{\rho})=0$, and consider the following cycle: (a) keeping the charge at the point $P$, displace the ions vertically into the configuration $\mathbf{w}(\boldsymbol{\rho})=\boldsymbol{\psi}(\boldsymbol{\rho})=\psi(\boldsymbol{\rho})\mathbf{e}_z$, according to 
\begin{equation}
\mathbf{w}(\boldsymbol{\rho})=\xi\psi(\boldsymbol{\rho})\mathbf{e}_z, 
\label{wroxiapz2}
\end{equation}
by increasing $\xi$ from 0 to 1; (b) keeping the ions at $\mathbf{w}(\boldsymbol{\rho})=\boldsymbol{\psi}(\boldsymbol{\rho})$, displace the charge vertically by a small $\Delta z$ to point $P'$ ($\boldsymbol{\rho}_0=0$,$z_0+\Delta z$); (c) fixing the charge at the point $P'$, reverse process (a), i.e., by reducing $\xi$ from 1 to 0  according to Eq.~(\ref{wroxiapz2}); (d) move the charge back to point $P$ to complete the cycle. 

The work per unit area required to change $\mathbf{w}(\boldsymbol{\rho})$ to $\mathbf{w}(\boldsymbol{\rho})+\Delta \mathbf{w}(\boldsymbol{\rho})$ is $-[c_{11}\mathbf{w}(\boldsymbol{\rho})+c_{12}\mathbf{E}_z(\boldsymbol{\rho},0)]\cdot \Delta \mathbf{w}(\boldsymbol{\rho})$, and thus total work expended on the ionic system for the configuration change is 
\begin{equation}
-\int [c_{11}\mathbf{w}(\boldsymbol{\rho})+c_{12}\mathbf{E}_z(\boldsymbol{\rho},0)]\cdot \Delta \mathbf{w}(\boldsymbol{\rho})d\boldsymbol{\rho}.  
\label{dewpaap1z}
\end{equation}

From Eq.~(\ref{wroxiapz2}) one has 
$\Delta\mathbf{w}(\boldsymbol{\rho})=\Delta\xi\psi(\boldsymbol{\rho})\mathbf{e}_z$.  Expanding $\mathbf{w}(\boldsymbol{\rho})$ and $\psi(\boldsymbol{\rho})$ as in-plane motion above, 
one finds $\mathbf{w}_{\mathbf{k}}$ in Eq.~(\ref{Ez0apz2}) is given by 
$\mathbf{w}_{\mathbf{k}}=\xi \boldsymbol{\psi}_{\mathbf{k}}$. 
Substituting these expansions in terms of  $\psi_{\mathbf{k}}$ together with Eq.~(\ref{Ez0apz2}) into the above expression (\ref{dewpaap1z}), and integrating over $\xi$ from 0 to 1, one obtains for the work expended during process (a), 
\begin{align}
W_a&=\sum_{\mathbf{k}}\left[\frac{1}{2}A\left(\frac{4\pi c_{12}c_{21}(\Lambda-k/2)}{1+4\pi c_{22}(\Lambda-k/2)}-c_{11}\right)\psi_{\mathbf{k}} \right.  \nonumber \\ &\qquad {} 
\left. +\frac{2\pi c_{12}e_{\nu}}{1+4\pi c_{22}(\Lambda-k/2)}e^{-k z_0}\right]\psi_{-\mathbf{k}}.  
\label{workpaap1z}
\end{align}

During process (b) the field acting on charge $e_{\nu}$ due to the ionic displacements is $\mathbf{E}_{z,w}(0,z_0)$, namely, the sum of the terms containing $\mathbf{w}_{\mathbf{k}}$ of Eq.~(\ref{Ezz0apz3}), with $\mathbf{w}_{\mathbf{k}}= \boldsymbol{\psi}_{\mathbf{k}}$, and the work expended  for this field of the ionic displacements is 
\begin{equation}
W_{b,w}=-2\pi e_{\nu}c_{21}\sum_{\mathbf{k}} \frac{k\psi_{\mathbf{k}}}{1+4\pi c_{22}(\Lambda-k/2)}e^{-k z_0}\Delta z.   
\label{workpbwap1z}
\end{equation}
The field acting on charge $e_{\nu}$ due to the charge itself is $\mathbf{E}_{z,e_{\nu}}(0,z_0)$, i.e., the sum of the terms containing $e_{\nu}$ of Eq.~(\ref{Ezz0apz3}), and  the work done for this field can be simply written as $W_{b,e_{\nu}}=-e_{\nu}E_{z,e_{\nu}}(0,z_0)\Delta z$.

Process (c) is the reverse of process (a) except for the displacement of charge $e_{\nu}$. Therefore upon reversing the sign of  expression (\ref{workpaap1z}) and multiplying its latter summands by a factor $e^{-k \Delta z}$, we readily find the work expended during the process,
\begin{align}
W_c&=-\sum_{\mathbf{k}}\left[\frac{1}{2}A\left(\frac{4\pi c_{12}c_{21}(\Lambda-k/2)}{1+4\pi c_{22}(\Lambda-k/2)}-c_{11}\right)\psi_{\mathbf{k}} \right. \nonumber \\ &\qquad {} 
\left. +\frac{2\pi c_{12}e_{\nu}}{1+4\pi c_{22}(\Lambda-k/2)}e^{-k(z_0+\Delta z)}\right]\psi_{-\mathbf{k}}.   
\label{workpcap1z}
\end{align}

During process (d) $\mathbf{w}(\boldsymbol{\rho})=0$, and according to  Eq.~(\ref{Ezz0apz3}) no field is associated with displacement $\mathbf{w}(\boldsymbol{\rho})$, and accordingly no work is needed for the displacement part,  $W_{d,w}=0$. 
For the field due to the point charge itself, the work done  is given by reversing  the sign of $W_{b,e_{\nu}}$, i.e.,  $W_{d,e_{\nu}}=e_{\nu}E_{z,e_{\nu}}(0,z_0)\Delta z$.  

As $W_a+W_{b,w}+W_{b,e_{\nu}}+W_c+W_{d,w}+W_{d,e_{\nu}}=0$, one finds 
\begin{align}
&2\pi e_{\nu}\left[c_{12}\sum_{\mathbf{k}}\frac{\psi_{-\mathbf{k}}e^{-k z_0}}{1+4\pi c_{22}(\Lambda-k/2)}(1-e^{-k \Delta z}) \right. \nonumber \\ &\qquad {} 
\left. - c_{21}\sum_{\mathbf{k}}\frac{k\psi_{\mathbf{k}}e^{-k z_0}}{1+4\pi c_{22}(\Lambda-k/2)}\Delta z\right]=0. 
\label{worksap2z}
\end{align}
Using $e^{-k \Delta z}=1-k \Delta z$, and changing the index of summation $\mathbf{k}$ to $-\mathbf{k}$ for the former summation, one readily finds $c_{12}=c_{21}$. 

Having $c_{12}=c_{21}$, now an areal energy density associated with the out-of-plane optical vibrations can be introduced,  
\begin{equation}
u_v=-\frac{1}{2}\left[c_{11}\mathbf{w}^2_z(\boldsymbol{\rho})+2c_{12}\mathbf{w}_z(\boldsymbol{\rho})\cdot\mathbf{E}_z(\boldsymbol{\rho},0)+c_{22}\mathbf{E}^2_z(\boldsymbol{\rho},0)\right],   
\label{endenuv}
\end{equation}
from which the lattice equations (\ref{eomw1zco}) and (\ref{bigP1zco}) can be rederived through $\ddot{w}_z=-\partial u_v/\partial w_z$, $\mathcal{P}_z=-\partial u_v/\partial E_z$ [$w_z=w_z(\boldsymbol{\rho})$, and $E_z=E_z(\boldsymbol{\rho},0)$]. 

\newpage


%

\newpage

\begin{table*}[htbp]
	\begin{center}
		\leavevmode
		\setlength{\tabcolsep}{8pt}
		\renewcommand\arraystretch{1.5}
		\caption{\label{table:1} Two groups of calculated lattice-dynamical  quantities associated with the in-plane optical vibrations in ML hBN, namely, the intrinsic oscillator frequency $\omega_0$, the 2D electronic susceptibility $\chi_e$ and static susceptibility $\chi_0$, the Born charge $e_B$, the static effective charge $e_a$, the effective spring force constant $K$, the force constant due to LFEs $K_e$, and the LO phonon group velocity $c_l$. In the upper (lower) row, the set of three quantities $\omega_0$, $\chi_e$, $\chi_0$ 
($\omega_0$, $\chi_e$, $e_B$) are obtained from first-principles calculations of Refs.\cite{Ferrabone:2011,Erba:2013} (Ref.\cite{Sohier:2017}), and then used to calculate the other quantities in this table with our expressions accounting for both EP and LFEs (see text).  }
		\begin{tabular}{c c c c c c c c}
			\hline
			\hline 
			$\omega_0$ (cm$^{-1}$)  & $\chi_e$ ($\AA$) & $\chi_0$ ($\AA$) & $e_B$ ($e$) & $e_a$ ($e$) &   $K$ (eV/$\AA^2$) & $K_e$ (eV/$\AA^2$) &  $c_l$ (km/s)   \\  \hline
			 1371$^a$  & 0.85 & 1.31 & 2.70 &  0.61  &  59.796 & 17.613 &  37.24    \\ 
			 1387$^b$  &  1.22  &  1.67  &  2.71  &  0.46  &  56.495 & 13.317 &  37.10  \\ 
			\hline
			\hline 
		\end{tabular}			
	\end{center}			     
 $^a$ First-principles perturbation result of Ref.\cite{Erba:2013}, equal to the experimental value \cite{Rokuta:1997} of ML hBN on substrate Ni and very close to DFPT value 1378 cm$^{-1}$ of Ref.\cite{Wirtz:2003}. \\ 
 $^b$ DFPT value from Ref.\cite{Sohier:2017}.      
\end{table*}

\newpage

\begin{table*}[htbp]
	\begin{center}
		\leavevmode
		\setlength{\tabcolsep}{5pt}
		\renewcommand\arraystretch{1.5}
		\caption{\label{table:2} Expressions for 2D electronic susceptibility $\chi_e$, 2D static susceptibility $\chi_0$, Born charge $e_B$, TO phonon frequency $\omega_t$, LO phonon dispersion $\omega_l(k)$, and LO phonon group velocity $c_l$ associated with the in-plane optical vibrations of ML hBN in the RIM and PIM, obtained with or without taking LFEs into account (see text). Below the expressions are the values of these physical quantities and also the specific quantitative dispersion $\omega_l(k)$, obtained with a set of first-principles values  $\chi_e=0.85~\AA$, $\chi_0=1.31~\AA$, $\omega_0$=1371 cm$^{-1}$ (put in the last row); values of $\chi_e$ and $\chi_0$ are in $\AA$, $e_B$ in $e$,  $\omega_t$ and $\omega_l$ in cm$^{-1}$ and $c_l$ in km/s. 
 }
		\begin{tabular}{p{0.6cm}<{\centering} p{1.3cm}<{\centering} p{1.9cm}<{\centering} p{2.2cm}<{\centering} p{1.4cm}<{\centering} p{1.5cm}<{\centering} p{3.3cm}<{\centering} p{2.2cm}<{\centering}}
			\hline
			\hline 
			\multicolumn{2}{c}{model}   & $\chi_e$  & $\chi_0$ & $e_B$ & $\omega_t$   &  $\omega_l(k)$ &  $c_l$  \\  \hline
	\multirow{4}{*}{RIM} & \multirow{2}{*}{no LFEs}   & 0  & $\frac{e_a^2}{sK}$ &  $e_a$  & $\sqrt{\frac{K}{\bar{m}}}$ & $\sqrt{\frac{K}{\bar{m}}+\frac{2\pi e_a^2k}{\bar{m}s}}$ & $\frac{\pi e_a^2}{s\sqrt{\bar{m}K}}$ \\ 
   &  & 0 & 0.017 & 0.61 & 1632  & $\sqrt{1632.4^2+837.4^2\tilde{k}}$ & 1.61    \\   \cline{2-8} 
	  & \multirow{2}{*}{LFEs}  &  0 & $\frac{e_a^2}{s(K-e_a^2Q_1)}$ &  $e_a$   &  $\sqrt{\frac{K-e_a^2Q_1}{\bar{m}}}$  &  $\sqrt{\frac{K-e_a^2Q_1}{\bar{m}}+\frac{2\pi e_a^2k}{\bar{m}s}}$  &  $\frac{\pi e_a^2}{s\sqrt{\bar{m}(K-e_a^2Q_1)}}$ \\ 
    &  & 0 & 0.018 & 0.61 & 1577  & $\sqrt{1577^2+837.4^2\tilde{k}}$ & 1.67    \\     \hline 
		\multirow{4}{*}{PIM}	& \multirow{2}{*}{no LFEs}   &  $\frac{\alpha_1+\alpha_2}{s}$  & $\frac{\alpha_1+\alpha_2}{s}+\frac{e_a^2}{sK}$ & $e_a$   &  $\sqrt{\frac{K}{\bar{m}}}$  &  $\sqrt{\frac{K}{\bar{m}}+\frac{2\pi e_a^2k}{\bar{m}[s+2\pi(\alpha_1+\alpha_2)k]}}$  &  $\frac{\pi e_a^2}{s\sqrt{\bar{m}K}}$ \\ 
 &  & [0.24,0.32]$^\dagger$ & [0.26,0.34]$^\dagger$ & 0.61 & 1632  & $\sqrt{1632.4^2+\frac{837.4^2\tilde{k}}{1+\gamma\tilde{k}}}^\oplus$ & 1.61    \\    \cline{2-8}  
            & \multirow{2}{*}{LFEs} & Eq.~(\ref{a22eB}) & Eq.~(\ref{chiwps0}) &     &  Eq.~(\ref{wto2a})  &  Eq.~(\ref{wlo2a2})   &  Eq.~(\ref{cl})     \\   
  &  & 0.85  & 1.31 & 2.70 & 1371  & $\sqrt{1371^2+\frac{3691.5^2\tilde{k}}{1+13.42\tilde{k}}}$ & 37.24    \\ 			\hline
		\hline 
		\end{tabular}			
	\end{center}		
$\tilde{k}$ is the normalized wavevector (dimensionless), $\tilde{k}=k/\frac{2\pi}{a}$. 
 \\   $^\dagger$ An interval (see text).  \quad  $^\oplus$  $\gamma$ is dimensionless, $3.8172 \leq \gamma \leq 5.1144$. 	\\  Dispersion relations given by the specific value-substituted expressions $\omega_l(k)$ are also shown in Fig.~\ref{fig2}. 
\end{table*}

\newpage

\begin{table*}[htbp]
	\begin{center}
		\leavevmode
		\setlength{\tabcolsep}{12pt}
		\renewcommand\arraystretch{1.5}
		\caption{\label{table:3} Comparison of intrinsic oscillator frequency $\omega_0'$, static effective charge $e_a'$, Born charge $e_B'$, force constant due to LFEs $K_e'$ and  effective force constant $K'$ associated with the out-of-plane optical modes in ML hBN calculated with the PIM including LFEs, for two values of frequency $\omega_z$ taken from previous first-principles calculations (Refs.\cite{Erba:2013,Topsakal:2009,Miyamoto:1995,Wirtz:2003}), one experimental $\omega_z$ value (Ref.\cite{Rokuta:1997}) and one low frequency $\omega_z$ of 405 cm$^{-1}$, using 2D electronic susceptibility $\chi_e'=0.151$ $\AA$ and static susceptibility $\chi_0'=0.164$ $\AA$ calculated from first principles (Ref.\cite{Ferrabone:2011}). }
		\begin{tabular}{c c c c c c}
			\hline
			\hline 
		$\omega_z$ (cm$^{-1}$)  & $\omega_0'$ (cm$^{-1}$)  & $e_a'$ ($e$) & $e_B'$ ($e$) & $K_e'$ (eV/$\AA^2$) &  $K'$ (eV/$\AA^2$)   \\  \hline
			836$^a$  & 802  &  1.26  &  -0.27   &  7.136  &  21.575   \\ 
			800$^b$  & 768  &  1.20  &  -0.25   &  6.535  &  19.758   \\ 
			734$^c$  & 704  &  1.11  &  -0.23   &  5.501  &  16.630   \\ 
			405$^d$  & 389  &  0.61  &  -0.13   &  1.675  &  5.063   \\ 
			\hline
			\hline 
		\end{tabular}			
	\end{center}			     
 $^a$ First-principles perturbation result of Ref.\cite{Erba:2013} and direct method result of Ref.\cite{Miyamoto:1995}. \quad  $^b$ DFPT value of Refs.\cite{Wirtz:2003,Topsakal:2009}.  
  \quad  $^c$ Experimental value of ML hBN on substrate Ni of Ref.\cite{Rokuta:1997}. \\ 
\quad  $^d$ This frequency yields a static effective charge 0.61$e$ equal to the $e_a$ value for in-plane motion.    
\end{table*}

\newpage

\begin{table*}[htbp]
	\begin{center}
		\leavevmode
		\setlength{\tabcolsep}{8pt}
		\renewcommand\arraystretch{1.5}
		\caption{\label{table:4} Expressions for 2D electronic susceptibility $\chi_e'$, 2D static susceptibility $\chi_0'$, Born charge $e_B'$, intrinsic oscillator frequency $\omega_0'$ and phonon frequency $\omega_z$ associated with the out-of-plane optical vibrations of ML hBN in the RIM and PIM, obtained with or without taking LFEs into account (see text). Below the expressions are the values of these physical quantities obtained with a set of first-principles values, $\chi_e'=0.151$ $\AA$, $\chi_0'=0.164$ $\AA$, $\omega_z=836$ cm$^{-1}$ (given in the last row); values of $\chi_e'$ and $\chi_0'$ are in $\AA$, $e_B'$ in $e$,  and $\omega_0'$ and $\omega_z$ in cm$^{-1}$. 
 }
		\begin{tabular}{p{0.6cm}<{\centering} p{1.3cm}<{\centering} p{1.8cm}<{\centering} p{2.2cm}<{\centering} p{1.3cm}<{\centering} p{1.9cm}<{\centering} p{3.3cm}<{\centering}}
			\hline
			\hline 
			\multicolumn{2}{c}{model}   & $\chi_e'$  & $\chi_0'$ & $e_B'$ & $\omega_0'$   &  $\omega_z$   \\  \hline
	\multirow{4}{*}{RIM} & \multirow{2}{*}{no LFEs}   & 0  & $\frac{e_a'^2}{sK'}$ &  $e_a'$  & $\sqrt{\frac{K'}{\bar{m}}}$ & $\infty$  \\ 
   &  & 0 & 0.196 & 1.26 & 981  & $\infty$     \\   \cline{2-7} 
	  & \multirow{2}{*}{LFEs}  &  0 & $\frac{e_a'^2}{s(K'+2e_a'^2Q_1)}$ &  $e_a'$   &  $\sqrt{\frac{K'+2e_a'^2Q_1}{\bar{m}}}$  &  $\infty$  \\ 
    &  & 0 & 0.076 & 1.26 & 1571  & $\infty$     \\     \hline 
		\multirow{4}{*}{PIM}	& \multirow{2}{*}{no LFEs}   &  $\frac{\alpha_1'+\alpha_2'}{s}$  & $\frac{\alpha_1'+\alpha_2'}{s}+\frac{e_a'^2}{sK'}$ & $e_a'$   &  $\sqrt{\frac{K'}{\bar{m}}}$  &  $\sqrt{\frac{K'}{\bar{m}}+\frac{e_a'^2}{\bar{m}(\alpha_1'+\alpha_2')}}$   \\ 
 &  & [0.36,0.48]$^\dagger$ & [0.55,0.67]$^\dagger$ & 1.26 & 981  & [1165,1220]$^\dagger$     \\    \cline{2-7}  
            & \multirow{2}{*}{LFEs} & Eq.~(\ref{c22z}) & Eq.~(\ref{chiwzs0}) &    &  Eq.~(\ref{c11z})  &  $\sqrt{\frac{K'+2e_a'e_B'Q_1}{\bar{m}}+\frac{e_B'^2}{\bar{m}s\chi_e'}}$        \\   
  &  & 0.151  & 0.164 & -0.27 & 802  & 836     \\ 		\hline
		\hline 
		\end{tabular}			
	\end{center}		
  $^\dagger$ An interval.  	
\end{table*}

\newpage

\begin{figure}

\caption
{(Color online) Schematic diagram of the honeycomb structure of monolayer hexagonal BN with the Bravais lattice basic vectors $\mathbf{a}_1$ and $\mathbf{a}_2$.  
}
\label{fig1}
\vspace*{15mm}

\caption
{(Color online)  Longitudinal optical (LO) phonon dispersion relations of monolayer hexagonal BN from the rigid ion model (RIM) and the polarizable ion model (PIM) with or without taking local field effects (LFEs) into account, which are described by the four specific value-substituted expressions for $\omega_l(k)$ in Table~\ref{table:2} (see text). For the PIM without LFEs $\gamma$ is taken to be 3.8172. Also shown is the LO phonon dispersion from Ref.\cite{Sohier:2017} obtained by  Sohier {\it et al}.. Here the wavevector is made dimensionless as in Ref.\cite{Sohier:2017}  with respect to $\lvert \Gamma-K\rvert$, the distance between the $\Gamma$ and $K$ points in the Brillouin zone. 
}
\label{fig2}
\vspace*{15mm}

\caption
{ 
(Color online)  Density of states (DOS) of the longitudinal optical (LO) phonon modes [expression (\ref{doslo1})]   (upper horizontal axis) and LO phonon dispersion [expression (\ref{wlo2a2})] (lower horizontal axis), calculated with the polarizable ion model (PIM) including the local field effects (LFEs) and using the first row of parameters of Table~\ref{table:1}. 
}
\label{fig3}
\vspace*{15mm}
\end{figure}

\newpage

\vspace*{20mm}

\begin{figure}
\includegraphics*[width=12cm]{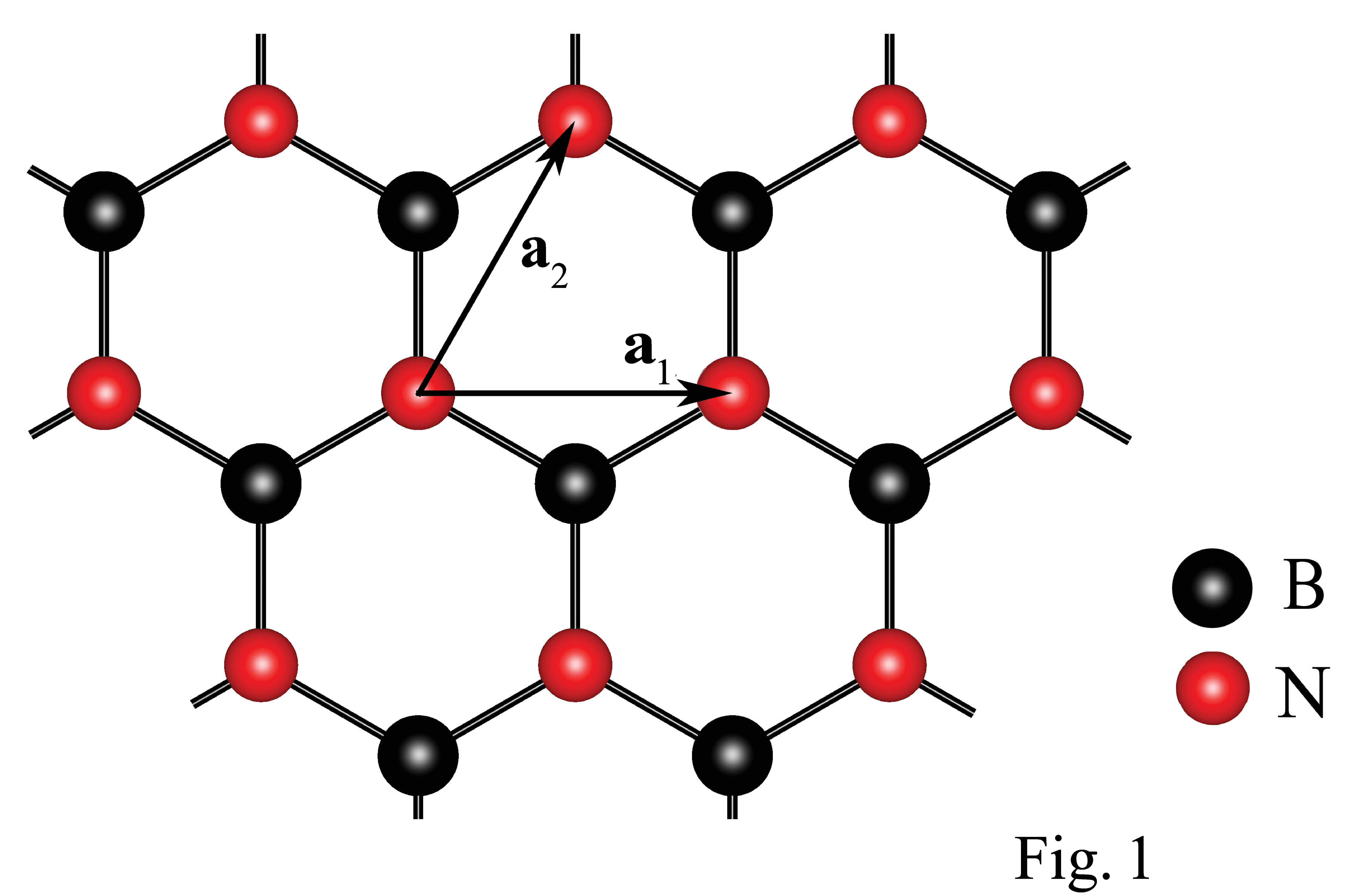}
\end{figure}

\newpage

\vspace*{20mm}

\begin{figure}
\includegraphics*[width=15cm]{fig2.eps}
\end{figure}

\newpage

\vspace*{20mm}

\begin{figure}
\includegraphics*[width=15cm]{fig3.eps}
\end{figure}

\end{document}